\documentclass[acmsmall]{acmart}

\AtBeginDocument{%
  \providecommand\BibTeX{{%
    \normalfont B\kern-0.5em{\scshape i\kern-0.25em b}\kern-0.8em\TeX}}}

\usepackage{graphicx}
\usepackage{booktabs} 
\usepackage{enumitem}
\usepackage{multirow}
\usepackage{cleveref}

\usepackage{diagbox}
\usepackage{subcaption}
\usepackage{amsmath}
\usepackage{dsfont}
\usepackage{marginnote}

\usepackage[group-separator={,}]{siunitx}
\usepackage{tabularx}

\definecolor{black}{rgb}{0.0, 0.0, 0.0}
\definecolor{navy}{rgb}{0.1, 0.1, 0.8}
\definecolor{ruby}{rgb}{0.6, 0.1, 0.3}
\definecolor{orange}{rgb}{1.0, 0.5, 0.0}

\usepackage{xspace} 
\newcommand{\header}[1]{{\noindent{\textbf{#1}}}}

\newcommand{\dataset}{{\sc covid2020}\xspace}
\newcommand{\covid}{{COVID-19}\xspace}

\begin{document}

\title{The Shapes of the Fourth Estate During the Pandemic: Profiling COVID-19 News Consumption in Eight Countries}

\author{Cai Yang}
\email{cai.yang@anu.edu.au}
\affiliation{%
  \institution{Australian National University}
  \city{Canberra}
  \state{ACT}
  \country{Australia}
}

\author{Lexing Xie}
\email{lexing.xie@anu.edu.au}
\affiliation{%
  \institution{Australian National University}
  \city{Canberra}
  \state{ACT}
  \country{Australia}
}

\author{Siqi Wu}
\authornote{Corresponding author}
\email{siqiwu@umich.edu}
\affiliation{%
  \institution{University of Michigan}
  \city{Ann Arbor}
  \state{Michigan}
  \country{USA}
}


\begin{abstract}
News media is often referred to as the Fourth Estate, a recognition of its political power. New understandings of how media shape political beliefs and influence collective behaviors are urgently needed in an era when public opinion polls do not necessarily reflect election results and users influence each other in real-time under algorithm-mediated content personalization. In this work, we measure not only the average but also the distribution of audience political leanings for different media across different countries. The methodological components of these new measurements include a high-fidelity \covid tweet dataset; high-precision user geolocation extraction; and user political leaning estimated from the within-country retweet networks involving local politicians. We focus on geolocated users from eight countries, profile user leaning distribution for each country, and analyze bridging users who have interactions across multiple countries. Except for France and Turkey, we observe consistent bi-modal user leaning distributions in the other six countries, and find that cross-country retweeting behaviors do not oscillate across the partisan divide. More importantly, this study contributes a new set of media bias estimates by averaging the leaning scores of users who share the URLs from media domains. Through two validations, we find that the new average audience leaning scores strongly correlate with existing media bias scores. Lastly, we profile the \covid news consumption by examining the audience leaning distribution for top media in each country, and for selected media across all countries. Those analyses help answer questions such as: Does center media {\it Reuters} have a more balanced audience base than partisan media {\it CNN} and {\it Fox News} in the US? Does far-right media {\it Breitbart} attract any left-leaning readers in any countries? Does {\it CNN} reach a more balanced audience base in the US than in UK and Spain? In sum, our data-driven methods allow us to study media that are not often collected in editor-curated media bias reporting, especially in non-English-speaking countries. We hope that such cross-country research would inform media outlets of their effectiveness and audience bases in different countries, inform non-government and research organizations about the country-specific media audience profiles, and inform individuals to reflect on our day-to-day media diet.
\end{abstract}

%
%


\setcopyright{acmlicensed}
\acmJournal{PACMHCI}
\acmYear{2023} 
\acmVolume{7} 
\acmNumber{CSCW2} 
\acmArticle{317} 
\acmMonth{10} 
\acmPrice{15.00}
\acmDOI{10.1145/3610108}

%

\maketitle

\section{Introduction}
\label{sec:intro}

News media is an important political force, as it is sometimes referred to as the Fourth Estate\footnote{\url{https://en.wikipedia.org/wiki/Fourth_Estate}} or fourth power, after the three estates -- the clergy, the nobility, and the commoners -- to reflect its power in advocacy and framing of political issues. Conventional knowledge about this fourth estate has been challenged in recent years, by unexpected election results~\cite{budak2019happened}, by alleged interference from foreign powers faraway~\cite{badawy2018analyzing}, by the widespread use of proprietary, algorithm-mediated content personalization~\cite{covington2016deep}, and by the fact that social media users influencing each other in real time~\cite{aral2021hype}. New understandings of how media reach and influence an inter-connected audience base around the world are very much called for, especially since much research has focused on the United States (e.g., its media, users, and political systems)~\cite{Budak2016FairAB,wu2021cross,Bakshy2015ExposureTI} and a small number of Western countries~\cite{fletcher2020polarized,lo2014common}.

This paper aims to address three prominent gaps on measuring contemporary media consumption. First, recent work on quantifying media bias -- via either editorial ratings~\cite{mbfc,allsides}, surveys~\cite{Budak2016FairAB,fletcher2020polarized}, or data-driven measurements~\cite{Bakshy2015ExposureTI,robertson2018auditing} -- all focuses on generating one single numerical score or ordinal label for each media, without explicitly taking into account the breadth of audience base or the diversity of published content. However, characterizing the distribution of audience on their political leaning or other attributes is important to reflect the diversity in both content providers and media consumers~\cite{Kwak2018WhatWR}, hence the metaphorical ``Shape of the Fourth Estate'' in the paper title. The second aspect is in measuring media outlets beyond the US and most-studied Western countries~\cite{lietz2014politicians,takikawa2017political,capozzi2021clandestino}, and designing automated methods and data-driven metrics that can generalize to a larger set of understudied media and countries. Quantitative profiles of media consumption in a diverse collection of countries can catalyze conceptual advances in characterizing media systems~\cite{hallin2004comparing,dobek2010comparative}, especially for societies outside the Western democratic system and culture. 
The third is the analysis of cross-country differences in audience profiles. Existing studies on media consumption either focus on in-depth analysis within a single country~\cite{Bakshy2015ExposureTI,lietz2014politicians,robertson2018auditing}, or a number of countries by separately accounting for their respective political systems~\cite{huszar2022algorithmic}. To the best of our knowledge, there has not been comprehensive political profiling of media audiences for the outlets across different countries. Overall, our work seeks to answer two questions:

\begin{enumerate}[label=\textbf{RQ\arabic*:}]
  \item For a given country, in terms of political leaning, what is the breadth of readership for the media outlets there? 
  \item For a given media, in terms of political leaning, how does its audience profile vary across different countries?
\end{enumerate}

To answer those questions, we collect one of the largest \covid tweet datasets from March to November 2020. We refer to this one billion tweet dataset as \dataset (\Cref{sec:dataset}). In the data processing pipeline, we first extract high-precision Twitter user locations from geotagged tweets and self-reported location strings (\Cref{ssec:method_geo}). We select eight countries where we can identify a sufficient number of politicians on the Twittersphere. We use the politicians' political position labels as seeds to estimate the user political leanings by applying a label spreading algorithm~\cite{zhou2003learning} over the user-user retweet networks (\Cref{ssec:method_leaning}). We profile the overall user leaning distributions in these eight countries, and link the observations to existing theories in comparative media literature~\cite{hallin2004comparing} (\Cref{ssec:result_ridge}). Next, we profile the group of bridging users who have cross-country retweets (\Cref{ssec:result_bridging}). We find consistent behavioral patterns -- if users mostly retweet one political side in a country, it is unlikely that they would turn to mostly retweeting the opposite side in another country, indicating that the political leaning scores estimated from the within-country retweet networks are comparable with each other.

We further derive a new media bias estimate by taking the average of user leaning scores from those who have shared URLs from this media (\Cref{ssec:result_biasscore}). We find these new estimates are highly correlated with existing media bias ratings by conducting two validations: (a) against six US-centric media bias scores~\cite{allsides,pewscores,Bakshy2015ExposureTI,Budak2016FairAB,robertson2018auditing}; (b) against a recent survey study for 12 different countries~\cite{fletcher2020polarized}. Using URL sharing as a signal of media consumption, this work contributes a series of nuanced profiles of audience leaning distributions for a specific media in a specific country (\Cref{ssec:result_consumption}). From a country-centric view, we show the country-level top consumed media and their audience leaning distributions. From a media-centric view, we show the different audience bases of one media across different countries. Those two views offer answers to many questions. Revisiting the questions we pose in the abstract, we find center media (e.g., {\it reuters}) tends to have a more balanced leaning distribution than partisan media (e.g., {\it cnn} and {\it foxnews}) in the US. As the partisan attitude of a media becomes stronger, or even extreme, fewer and fewer users from the opposite side would choose to consume its news (e.g., {\it breitbart}). Moreover, even if a media has an imbalanced audience base in one country (e.g., {\it cnn} in the US), it may still have a more balanced audience base in another country (e.g., {\it cnn} in Spain). All of those questions cannot be readily answered if one opts to disregard the audience leaning distribution in different countries.

Overall, our work presents a data-driven profiling of country-specific audience leaning distributions, which provides new insights to media outlets, activist groups, and media watchdog organizations, as well as to individuals for reflecting on and choosing our day-to-day media diet. We hope our methodology and observations serve as a basis for further inquiry, and they will shed light on understanding political behavior and media landscape in countries that do not often appear in computational social science writings, such as the Global South countries. 

In sum, the main contributions of this work include:

\begin{itemize}[leftmargin=*]
  \item A new, large, and longitudinal tweet dataset \dataset.\footnote{Code and data are publicly available at \url{https://github.com/computationalmedia/media_landscape}} 
  \item A new set of media bias estimates (surrogated by the average audience leaning score from users who have shared URLs from this media) that are shown to be highly correlated with existing editorial and survey-based media bias scoring. 
  \item A new analysis of bridging users across countries, showing that cross-country retweeting behaviors do not oscillate across the partisan divide.
  \item A fine-grained profiling of audience political leaning distribution for a given media in a given country, which offers new insights to many comparative questions.
\end{itemize}

\section{Related Work}
\label{sec:related}

\subsection{Measuring Media Biases}

Having a reliable estimate of political media bias is important to scholars in political science, communication, and journalism. There have been three lines of approaches to measuring media bias. The first is via expert rating on media content. Several independent websites, such as~\citet{adfontes}, \citet{allsides}, \citet{mbfc}, all rely on a small editorial team to judge the content published by media and produce their own media bias reports. While the bias manifested in media content is seemingly a reasonable measure, the operation of expert rating makes the annotation process non-transparent, hard to scale, and subject to the biases from the human annotators. 
Second, there is a rich literature on automated identification of media bias, to name a few, text analysis based on the published news articles~\cite{d2022machine}, user analysis based on the interactions over multiple communities~\cite{waller2021quantifying}, network flow analysis based on media's corporate contribution dollars to political parties~\cite{gentzkow2010drives}, and content analysis based on the exposure of political actors~\cite{kim2022measuring} or biased news coverage~\cite{vallone1985hostile}. For more discussion on this topic, we refer the reader to a recent survey~\cite{hamborg2019automated}.
The third line is using the leaning of a media's audience as a surrogate of media bias. This is under the general observation of selective exposure~\cite{stroud2010polarization}. Many studies have chosen this approach to estimate media bias~\cite{pewscores, robertson2018auditing, Bakshy2015ExposureTI, Budak2016FairAB}. One challenge here is how to obtain user leaning. In previous research, \citet{robertson2018auditing} linked Twitter users to US voter registration records, \citet{Bakshy2015ExposureTI} used self-reported political affiliation from Facebook US users, Pew Research Center~\cite{pewscores} surveyed partisan US internet users about their media habits. Once obtaining the user leaning, for one media, they usually computed the relative difference of Democrats and Republicans (or liberals and conservatives) who had interacted with this media as a metric of media slant. 

In this work, we opt for the last approach. Differing from the aforementioned research~\cite{robertson2018auditing,Bakshy2015ExposureTI,pewscores} that leverages offline signals (e.g., voting records, demographics surveys), we use a computational method to infer user political leaning. We first identify a set of real-world politicians on Twitter, and use their political position labels as seeds to run a label spreading algorithm~\cite{zhou2003learning} on the within-country user-user interaction network. This method has proved to work remarkably well in both our experiments and prior studies~\cite{stewart2017drawing,lee2022whose}. Furthermore, unlike most research~\cite{Bakshy2015ExposureTI,Budak2016FairAB,robertson2018auditing} that only reports the average of audience leaning scores, we profile the distributions of audience leaning. This allows us to measure the variance of media audience. Even if two media have the same political bias rating, they might have a huge difference in whom reads news from them. 
Many examples are given in~\Cref{fig:top15-domain}. For instance, while {\it politico} and {\it cbsnews} are both rated center-left media, the audience of {\it politico} are mostly left-leaning, but the audience of {\it cbsnews} consist of a diverse set of left-leaning and right-leaning users. See the detailed analysis in~\Cref{ssec:result_consumption}.

\subsection{Cross-Country Comparative Media Studies}

The research of media consumption receives unequal attention geographically with many studies only focusing on the US~\cite{Bakshy2015ExposureTI,Budak2016FairAB,pewscores,wu2021cross,an2014partisan,eady2019many}. This is in part because US users have shown high degree of polarization in many online and offline behaviors~\cite{druckman2021affective}, and in part because the two-party political system operating in US, which naturally fits into the left-right or liberal-conservative political spectrum. For a comprehensive picture of media consumption in multiple countries, we refer to the Reuters Institute's yearly digital news reports~\cite{newman2022reuters}.

Cross-country comparative media research is important yet challenging. As~\citet{livingstone2003challenges} highlights, ``folk wisdom cautions against comparing apples and oranges'' (p. 480). The problem is rooted in the gaps in data collected following different procedures, measures estimated over different populations, and concepts drifted across different countries. One canonical example is the contrasting attitudes toward abortion rights over the world. In the US, abortion attitudes are highly polarized by political ideology and abortion policies vary by state; while in many European countries, abortion is generally permitted. This example illustrates that abortion attitudes may be an effective signal to identify discordant groups in the US, but will become ineffective if one wants to apply it in European countries.

There is a rich line of comparative research; however, we argue that some are not really cross-country analyses because the within-country observations should be calibrated for cross-country comparisons. For example, \citet{huszar2022algorithmic} audited the algorithmic amplification of political content on Twitter in seven countries. They measured the attitudes toward major political parties in each country -- two parties in the US, four in the UK, five in Japan, six in Germany, and seven in France. While the measures in each individual country are valid, there is no warrant for the comparability across countries. In order to conduct rigorous comparative research, one must construct a common measurement axis~\cite{lo2014common}. For example, \citet{fletcher2021many} used surveys to directly ask panelists from different countries to label their political leaning on a left-right spectrum. While a survey is an effective way to obtain self-reported subjective labels, it greatly limits the number of users one can study and generally has a much higher cost. 

In this work, we scrutinize the cross-country measures before analysis. We identify groups of bridging users who have interactions in multiple countries. Except for France, we do not find any systematic shifts in the political leaning users estimated from the within-country retweet networks. This assures the comparability of the user political leaning estimates in different countries. See the detailed analysis in~\Cref{ssec:result_bridging}.

\section{A new {\sc covid2020} tweet dataset}
\label{sec:dataset}

We use the Twitter filtered streaming API to construct a new, high-volume \covid tweet dataset. Our initial investigation in early 2020 shows that the daily volume of \covid tweets exceeds Twitter API limit by a large margin. For example, about 25M \covid tweets were posted each day in March 2020. This is significantly over the Twitter API limit of 1\% of total tweets, which is roughly 4M for a day.\footnote{Researchers have found that the Twitter filter streaming API provides no more than 50 tweets per second~\cite{wu2020variation}, though the Twitter API limit is more commonly known as no more than 1\% of the entire tweet volume for any time intervals.} If we just use the Twitter API naively, more than 80\% of relevant tweets would be missing. A better data collection strategy is needed to increase the data coverage and recall. We thus adopt a crawling strategy proposed by~\citet{wu2020variation}, which consists of three steps: (a) partitioning the whole data stream into several sub-streams by the tracked keywords and languages; (b) estimating the sampling rate for each sub-stream using the rate-limit messages provided by Twitter; (c) taking the union of all sub-streams. This strategy is shown to significantly reduce data loss, and thus can result in a high-fidelity dataset for high-volume, real-time tweet stream under Twitter API's limits~\cite{wu2020variation}.

We ran the data crawler from March 2020 to November 2020, covering eight months in the first year of the pandemic. The list of tracked \covid keywords is obtained from~\citet{chen2020tracking} (see \Cref{app:keywords}). Those keywords include not only generic terms such as ``corona virus'', ``covid'', but also non-pharmaceutical interventions such as ``lockdown'', ``n95'', and ``social distancing''. Note that we do not periodically update the tracked keyword list to account for emerging \covid topics such as ``vaccine''. However, vaccination had not been rolled out in the US until December 2020, and in most other countries, not until 2021.\footnote{\url{https://ourworldindata.org/covid-vaccinations}} We thus believe our keyword list reflects the public interest in \covid during the data collection period.

Our sub-stream configurations are shown in \Cref{app:keywords}. The estimated sampling rates of different sub-streams range from 95\% to 100\%, meaning that we have a very high recall for all \covid posts on the Twittersphere. To reduce the computational load, we processed one week's data in every two weeks. The resulting dataset contains 18 calendar weeks over an eight-month period (Mar-Nov, or week 13 to week 47 in 2020). We experienced several server glitches during data collection, and lost data for two entire weeks (week 27, 29) and another four days (in week 17, 31). In total, we obtained \num{999040035} \covid tweets posted by \num{62687121} users. We refer to this dataset as \dataset.\footnote{To comply with Twitter’s content redistribution policy, we only release the tweet IDs of the collected tweets. We cannot directly release our processed data because user-level information (e.g., extracted geolocation and estimated political leaning) is considered sensitive personal information by Twitter and by~\citet{gdpr2016regulation}.}

\begin{figure*}[tbp]
    \centering
    \includegraphics[width=.8\columnwidth]{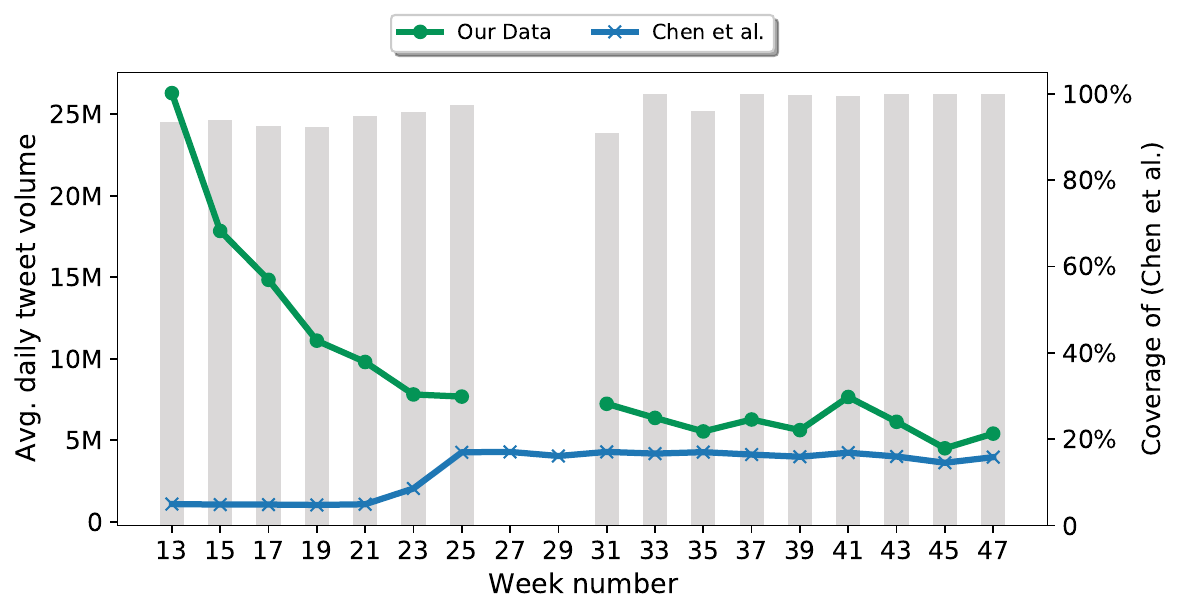}
    \caption{Comparison between our \dataset dataset and \citet{chen2020tracking}. Left y-axis: \dataset (green line) is much larger than \citet{chen2020tracking} (blue line) throughout time. Right y-axis: most tweets (on average 96.6\%) from \citet{chen2020tracking} are also included in \dataset (gray bar).}
    \label{fig:data_comparison}
\end{figure*}

The temporal data volume of the \dataset dataset is shown in \Cref{fig:data_comparison} (green line). The x-axis is the week number. The left y-axis shows the average number of daily tweets in the selected week. The volume of \dataset dataset starts from more than 25M tweets per day in March, and then drops to about 5M tweets per day in November. The decline may be attributed to user fatigue. According to a survey from Pew Research Center~\cite{pewfatigue}, 71\% of Americans expressed willingness to take breaks from \covid news. Another possible reason is the surfacing of new and uncollected topics. For instance, the discussions about \covid shift from lockdown to vaccine over time.

\header{Significance and bias of the \dataset dataset.}
On April 29, 2020, Twitter announced its own \covid data stream endpoint and restricted access to those who applied before October 2020\footnote{\url{https://twittercommunity.com/t/new-covid-19-stream-endpoint-available-in-twitter-developer-labs/135540/5}}. Unfortunately, despite that our data collection began in March 2020, we only became aware of this data source after the application deadline. Thus we could not provide a direct comparison with Twitter's \covid data stream. From the documentation page archived by the Wayback Machine,\footnote{\url{https://web.archive.org/web/20221025172340/https://developer.twitter.com/en/docs/twitter-api/tweets/covid-19-stream/filtering-rules}} we find that Twitter's tracked \covid keyword list is much larger than ours because Twitter includes many non-English terms from countries where people usually refer to ``corona virus'' in their corresponding native languages, for instance, China, Japan, Korea, India, Indonesia, and the Arab world. This indicates that our \dataset dataset is not a representative sample for public \covid discourse and Twitter users in non-English-speaking countries and we should exclude those countries from our analysis. For countries where people still primarily use ``corona virus'', ``covid'', and/or their stems to refer to \covid, we expect the \dataset dataset to have a high coverage for \covid tweets posted there. For instance, people say ``le coronavirus'', or ``le virus corona'', or ``le (virus) Covid dix-neuf'' in French. All of those keyword variants are captured by our data collector.

We also compare the \dataset dataset with a widely cited \covid tweet dataset constructed by~\citet{chen2020tracking} (\Cref{fig:data_comparison} blue line). We find that \dataset is significantly larger throughout the data collection period, though the volume difference decreases as time passes. For example, in week 13 (03/23 - 03/29), \citet{chen2020tracking} had only 7.6M tweets while we collected 184M tweets (24 times larger). Note that the observation of decreasing volume of \covid tweets is only made possible by our data collection strategy with a high-sampling rate (green line). Reading the daily volumes from a single Twitter data stream would lead to the opposite and incorrect conclusion -- i.e., rising \covid tweet volume (blue line). On the right y-axis, we show the fraction of tweets from~\citet{chen2020tracking} that also appear in \dataset for every week (gray bar). On average, 96.6\% of tweets from~\citet{chen2020tracking} are included in \dataset. To the best of our knowledge, \dataset is the largest publicly available \covid tweet dataset for the same period.

\section{Extracting User Geolocation, Political Leaning, and Shared Media Domains}

In this section, we extract attributes from tweets and Twitter users that are necessary for profiling \covid news consumption in different countries, namely high-precision user geolocation (\Cref{ssec:method_geo}), user political leaning in the left-right spectrum (\Cref{ssec:method_leaning}), shared URLs and media domains (\Cref{ssec:method_url}).

\subsection{Extracting Geolocation from Geotagged Tweets and User Profiles
}
\label{ssec:method_geo}

On Twitter, we can extract the location information from two places: (a) geotag embedded in each tweet, which has high precision but is known to have low coverage and (b) location field in the user profile, which has higher coverage but can be very noisy.

\begin{itemize}[leftmargin=*]
  \item \textbf{Geotagged users.} 0.85\% of all tweets in \dataset are attached with geotags, comparable to a prior study~\cite{pfeffer2023just}. A geotagged tweet includes a (city, state, country) place tag. We use the country tag as the surrogate of user country. To aggregate geotagged tweets into geotagged users, we assign all the extracted locations to a user, which builds the user mobility traces during the \covid pandemic. For simplicity, we filter out all users posting geotagged tweets from multiple countries -- a reasonable criterion due to the significantly reduced global mobility in our data collection period. With this method, we identify \num{1958826} geotagged users.
  \item \textbf{Geoparsed users.} Because most users do not post geotagged tweets, we also parse the self-reported location strings from the user profiles. We use the \textit{Simplemaps}\footnote{\url{https://simplemaps.com/data/world-cities}} data to construct an exhaustive list of all possible combinations of cities, states, and countries (in both full names and abbreviations). We check the location strings against the curated list, and we only include users with one exact string-matching location. Restricting geoparsing to exact matches improves precision, and eliminates free-form imaginary locations such as {\it neverland} or {\it Mars} that are commonplace on Twitter. We identify \num{20975218} geoparsed users.
\end{itemize}

\header{Evaluation.}
At the country level, there are \num{1077887} overlapping users from the two aforementioned methods. We consider the locations obtained by geotagged tweets as ground-truth and then evaluate the precision of parsing self-reported location strings on the overlapping users. We find that geoparsing correctly identifies the locations for 93\% of overlapping users. The high precision shows that it is possible to take advantage of both methods to extract user locations on Twitter. We hence merge the results of two methods, with mismatched users assigned locations from geotagging. In total, there are \num{21097109} users with one unique geolocation, accounting for 33.6\% users in \dataset. They have posted \num{397943516} tweets, accounting for 39.8\% tweets in \dataset. 

There are several third-party libraries for extracting geolocations from social media data~\cite{middleton2018location}, for example, OpenStreetMap and Google Geocoder API. We experimented with those libraries in our pilot test. However, we find that those libraries often operate based on a quota system and thus are unable to process millions of requests in a reasonable amount of time. We also evaluated several machine-learning-based geoparsing libraries (e.g., Mordecai) that trained on location texts, but found their performances not satisfying on the free-text Twitter profile location strings. 

\begin{figure*}[tbp]
    \centering
    \subcaptionbox{\centering \label{subfig:user-selection}}[.51\columnwidth][c]{%
    \includegraphics[width=.49\columnwidth]{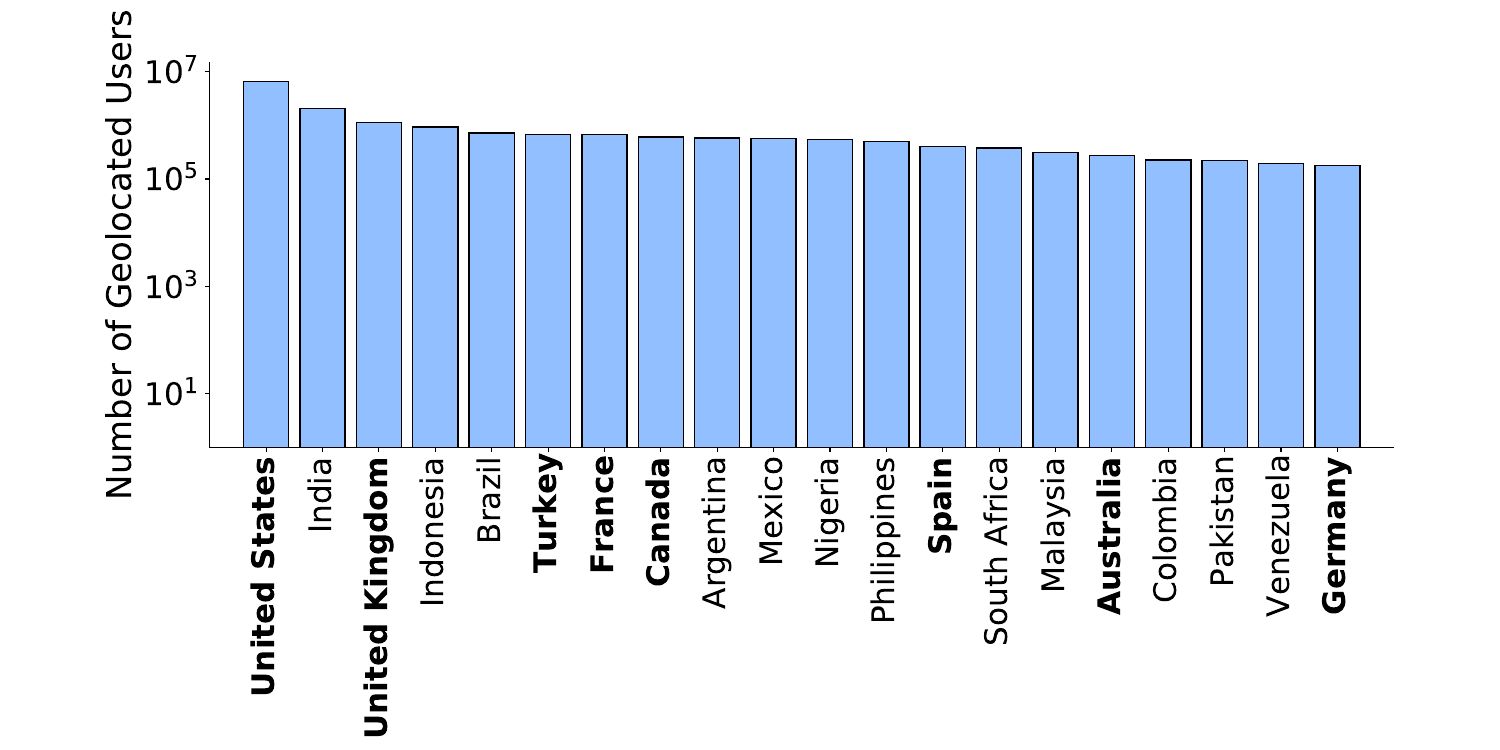}}
    \subcaptionbox{\centering \label{subfig:top2-lang}}[.48\columnwidth][c]{%
    \includegraphics[width=.48\columnwidth]{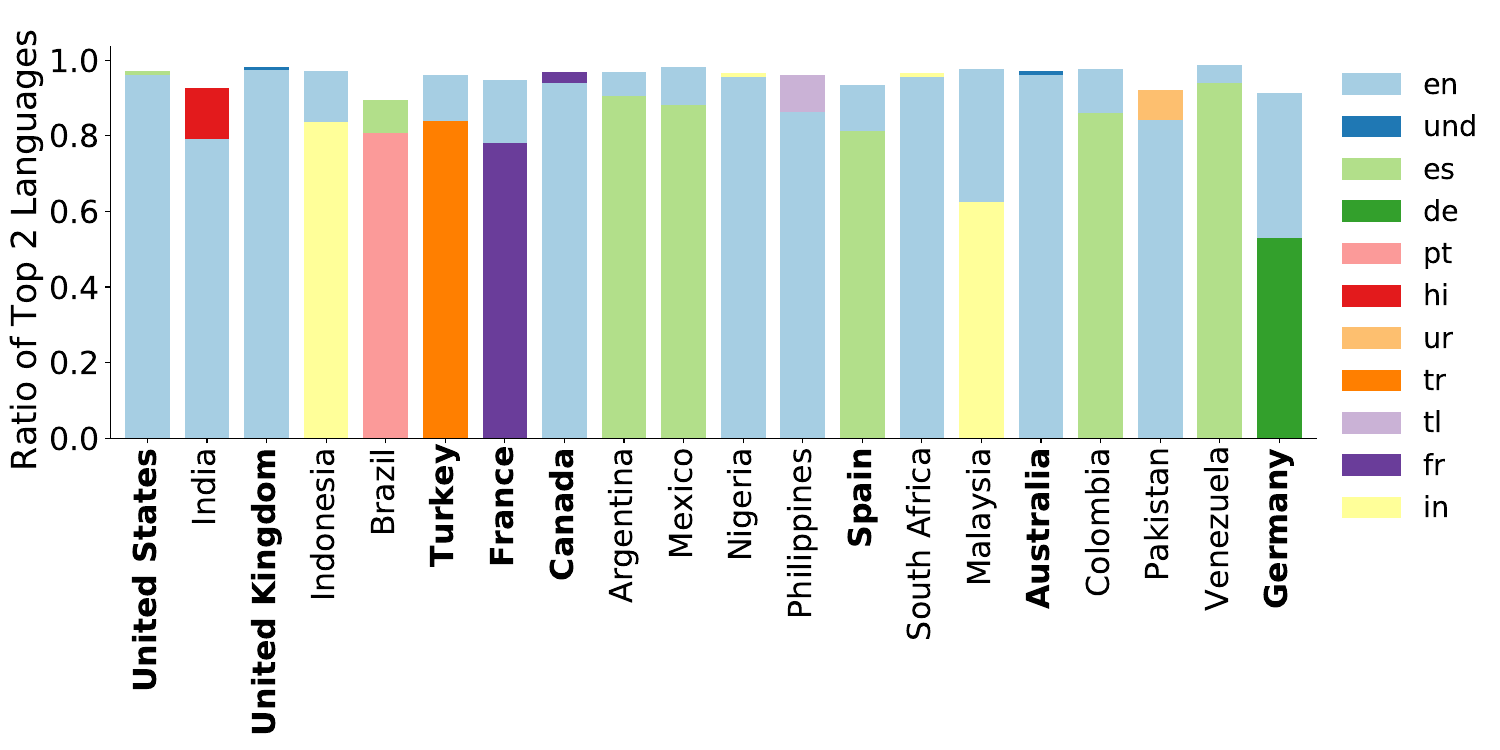} }
    \caption{(a) Top 20 countries with the most geolocated Twitter users. (b) The fractions of the two most used languages in the top 20 countries. Bold texts mark the final set of eight selected countries.
    }
\end{figure*}

\header{Geolocated users broken down by country.}
\Cref{subfig:user-selection} shows the top 20 countries with the most geolocated users. US has the most users (\num{6638437}) while Germany has the least (\num{178571}). The country ranking of geolocated Twitter users is on par with previous work~\cite{lamsal2022twitter}. 

\header{Top tweet languages in each country.}
We perform another sanity-check of geolocated tweets by examining the language distribution in the top 20 countries.
Language information is extracted from the ``\texttt{lang}'' field in the tweet object. It can be one of the BCP 47 language codes or \texttt{und} if the language is not recognized by Twitter's language classifier.
\Cref{subfig:top2-lang} shows the stacked bar plot of the two most used languages in each country (to avoid the long tail of languages cluttering the legend). The y-axis shows the ratio of tweets written in the marked language. We find that a large fraction of tweets are posted using the native or official languages in non-English speaking countries. For instance, Spanish (\texttt{es}) is the most used language in Argentina, Mexico, Spain, Colombia and Venezuela; Portuguese (\texttt{pt}) is most used in Brazil; Turkish (\texttt{tr}) in Turkey; and French (\texttt{fr}) in France. This assures us that the \dataset dataset covers global Twitter users who are representative of their residence countries, and not obviously biased toward English-speaking users.

\subsection{Estimating User Political Leaning from Within-Country User-User Retweet Networks}
\label{ssec:method_leaning}

Ideally, there is
a universal axis to measure the political polarization in different countries, but we find it an intrinsically difficult task due to the multifaceted nature of politics-diverse political systems. 
Commonly seen political facets include but are not limited to 
ideology (liberal vs. conservative), political leaning (left vs. right-leaning), party affiliation (Democratic vs. Republican), value (collectivism vs. individualism), and stance toward contested issues (pro vs. anti-abortion). After careful consideration, we find that some of those dimensions cannot be applied 
uniformly
at the global level. Firstly, the religious and social values, and issue stances have huge variations and sometimes are even misaligned in ideological groups across countries~\cite{pewvalue}. For example, US liberals support abortion while US conservatives oppose it; however, abortion rights are generally supported in Canada and Europe regardless of political ideology~\cite{politico2022}. If we use pro or anti-abortion stances to classify users, we would misclassify Canadian conservatives as liberals. Secondly, the political dividing issues vary significantly across countries. For example, in the US, contested issues include abortion, gun policy, racial justice, etc.~\cite {lee2022whose}. But in most European countries, debates between political groups primarily focus on taxation, welfare, and immigration~\cite{capozzi2021clandestino}. 
Issues in Asian and African countries are also less studied or documented.

For many countries, Wikipedia maintains a list of political parties and their political positions in a five-point Likert scale: left-wing, center-left, center, center-right, right-wing, which we collapse into three groups: left, center, right.\footnote{An example Wikipedia page of political parties in Canada: \url{https://en.wikipedia.org/wiki/List_of_federal_political_parties_in_Canada\#Current_parties}} We thus decide to use the left or right-wing party labels provided by Wikipedia to estimate the user political leaning (left or right-leaning) from the within-country user-user retweet network involving local politicians. The term ``within-country'' means that a user's predicted leaning would only be affected by how s/he interacts with other users and politicians from the same country. Researchers have found highly segregated behaviors of politicians with opposing party affiliation~\cite{gentzkow2019measuring,eady2020news,golovchenko2020cross}, as well as deepening polarization in the mass public ~\cite{lelkes2016mass,finkel2020political}. One can expect what partisan information a user proactively chooses to share on social media indicates the alignment of the user's own political leaning with the shared partisan. 

Note that the left and right-wing party labels are umbrella terms covering many social and economic issues. It is possible that two left-wing parties in different countries do not share a common belief over all issues~\cite{pewvalue}. Instead of dealing with nuanced differences directly, we rely on the crowd wisdom of Wikipedia contributors, who probably have more local political knowledge. It is also possible that assessments of left or right-wing party labels on Wikipedia do not follow the same criteria across countries. In other words, a left-wing party in one country may be perceived as right-wing by Wikipedia contributors in another country. To mitigate this concern, we empirically measure the cross-country retweet behavior on the set of bridging users in~\Cref{ssec:result_bridging}. As a preview, for users who retweet others in multiple countries, we do not find many people who mostly retweet one political side in one country, but turn to mostly retweeting the opposite side in another country. 

Specifically, our procedure contains three steps (details in the following subsections). Firstly, we collect the list of political parties in each country. We search for the party members servicing in the year 2020 and their Twitter accounts. Next, we construct the user-user retweet network for each country. Finally, using local politicians as the seed nodes, we apply the label spreading algorithm~\cite{zhou2003learning} to estimate the user's political leaning in the left-right spectrum. We remark on three important points for this procedure. First, the use of label spreading algorithm is based on the assumption of network homophily. Prior research has shown that the retweet network is of high homophily~\cite{conover2011political,lietz2014politicians} and can be used to identify political groups on Twitter~\cite{lietz2014politicians,barbera2015birds,lee2022whose}. Second, the political leaning score that we estimate in this work does not quantify the extent of leaning left or right. It is possible that a user who disproportionately retweets one political side has only a mild view and a user who sporadically retweets both sides has a more extreme view. For the problem of detecting online extremism, we refer the reader to~\cite{govers2023down}. Third, the score neither represents probability because it is not yet calibrated. One can calibrate the predicted scores to real probability scores by obtaining a human-annotated subsample and then applying the Platt scaling technique~\cite{platt1999probabilistic}. The best way to interpret the estimated political leaning scores is that how confident we are in asserting a user is left or right-leaning by measuring their relative frequency of broadcasting information from the left and right-leaning groups. 

\subsubsection{Collecting politicians' Twitter accounts}
\label{sssec:method_politician_acc}
Following~\cite{huszar2022algorithmic}, we use Wikidata's Query Service to identify politicians and their Twitter accounts. For a political party in a country, the Wikidata database lists its political position and party members, whose Twitter accounts are also indexed in Wikidata. For the top 20 countries with the most geolocated users, we collect the list of legislators and governors (or other equivalent roles). We require that they hold an active position in the year 2020. We also include all the election candidates from the 2020 United States presidential election.\footnote{\url{https://en.wikipedia.org/wiki/2020_Democratic_Party_presidential_candidates} and \url{https://en.wikipedia.org/wiki/2020_Republican_Party_presidential_primaries}} Not all politicians actively advocate their views on Twitter and the usage of Twitter for political activism varies significantly by countries. We compute the coverage of politicians with identified Twitter accounts and filter out countries where the coverage is lower than 40\%. This excludes 12 countries (India, Indonesia, Brazil, Argentina, Mexico, Nigeria, Philippines, South Africa, Malaysia, Colombia, Pakistan, and Venezuela) from our study. We are left with the eight countries shown in~\Cref{table:countries} (United States, United Kingdom, Canada, Australia, Spain, France, Germany, and Turkey).

\begin{table*}[t]
    \centering
    \resizebox{\textwidth}{!}{
    \begin{tabular}{rrrrrr}
    \toprule
    & \#politicians & \#w/ Twitter accounts & \%coverage & size of retweet network & \%coverage of geo. users \\ 
    \midrule
    United States & \num{609} & \num{593} & 97.4\% & \num{946436} & 14.3\% \\ 
    United Kingdom & \num{1456} & \num{636} & 43.7\% & \num{98061} & 8.7\% \\ 
    Canada & \num{454} & \num{192} & 42.3\% & \num{40103} & 6.7\% \\ 
    Australia & \num{235} & \num{176} & 74.9\% & \num{23838} & 8.7\% \\ 
    Spain & \num{635} & \num{530} & 83.5\% & \num{46122} & 11.4\% \\ 
    France & \num{943} & \num{550} & 58.3\% & \num{40576} & 6.1\% \\ 
    Germany & \num{752} & \num{574} & 76.3\% & \num{14508} & 8.1\% \\ 
    Turkey & \num{681} & \num{476} & 69.9\% & \num{27809} & 4.1\% \\ 
    \bottomrule 
    \end{tabular}
    }
    \caption{The number of politicians, number and coverage of politicians with Twitter accounts, size of within-country user-user retweet network, and coverage of geolocated users (i.e., geolocated users in the retweet network divided by all geolocated users) in the eight selected countries.}
    \label{table:countries}
\end{table*}

\subsubsection{Constructing the within-country user-user retweet network}
A simple retweet (without comment) disseminates the original post to the retweeter's followers, showing support from retweeter to retweetee.\footnote{We exclude retweets with comments (also known as ``quote tweet'' on Twitter) because the added comment may manifest disagreement or a negative attitude.} To this end, we construct the user-user retweet network for the geolocated users in each of the eight selected countries. The result is an undirected network. Each node is a geolocated user, and each edge indicates at least one retweet between the two users. We use the number of retweets between the two users as the edge weight. To remove insignificant edges in the network, we run the disparity filtering algorithm~\cite{Serrano2009ExtractingTM} to extract the network backbone. We set the significance threshold to be $0.05$ as suggested~\cite{Serrano2009ExtractingTM}. We make a minor modification to the filtering process: if an edge is statistically insignificant but connects nodes that are both seed users, we do not discard it to ensure better connectivity to seed users (i.e., politicians in this study). This modification is shown to help the label spreading step described later. As shown in~\Cref{table:countries}, US has the largest user-user retweet network and the highest coverage of geolocated users.

\subsubsection{Propagating labels from politicians to general users}
We apply the label spreading algorithm~\cite{zhou2003learning} on the extracted retweet network to infer the user's political leaning in each country. Local politicians are used as the seed nodes. An important hyperparameter is $\alpha$, which controls the trade-off between preserving original information in the focal node and receiving new information from the neighboring nodes. We use 10-fold cross-validation to do a line search between $0$ to $1$ and find the optimal $\alpha$ in each country respectively. Upon convergence, label spreading returns a score between 0 and 1, which is then rescaled into $[-1, 1]$. We interpret the rescaled score as the predicted user political leaning. A score closer to $-1$ ($1$) means this user disproportionately retweets information from other left-leaning (right-leaning) users. A score closer to 0 means this user interacts with people from both partisans. We use 10-fold cross-validation to measure the prediction results on the seeded politicians. The accuracy is consistently high ($0.90-0.98$) over all countries, with the US having an accuracy of 0.96 and France having the lowest accuracy of 0.90.

\header{Additional evaluation for the US users.} One drawback of the above performance evaluation is that the sizes of seed nodes (\#local politicians) are rather small, which may cause spuriously promising results. Because our \covid data collection period is leading up to the 2020 US presidential election, we find that many tweets in \dataset also contain hashtags related to the US election. This enables us to adopt a hashtag-based method to predict the leaning of US users by measuring the stances toward election candidates from the two major US political parties.

Specifically, we collect political dividing hashtags from previous work~\cite{Jiang2020PoliticalPD,wu2021cross}, for example, \textit{\#bidenharris2020}, \textit{\#voteblue} for left-leaning hashtags; \textit{\#maga}, \textit{\#trump2020} for right-leaning hashtags. We obtain 314 left-leaning and 246 right-leaning hashtags. The full list is available in \Cref{app:politics-hashtags}. For each user, we calculate a score by $\frac{R-L}{R+L}$, where $R$ and $L$ are the numbers of right- and left-leaning hashtags this user has posted. To reduce uncertainty, we only consider users with a score less than $-0.9$ (higher than $0.9$) as left-leaning (right-leaning) users. In total, we identify \num{195719} users (\num{83382} left, \num{112337} right). We then repeat the label spreading step on the US retweet work by using the \num{195719} users as seeds.
By this, we are able to estimate the political leaning for \num{1023093} US users, ${\sim}76K$ (8\%) more than using politicians as seeds. 

Comparing the two versions of prediction results (politician seeds vs. hashtag seeds), there are \num{600146} common users and 93.4\% of them have the same predicted political leaning. Although we still do not know about the ground-truth labels for those users,\footnote{The ground-truth political leaning has to be obtained by directly asking the users, e.g., via survey or interview, but not by any means of inference.} the high agreement rate between two different manners demonstrates the effectiveness of our overall data processing pipeline. For consistency, we use the politician-seeded predictions in the follow-up analysis.

\subsection{Extracting Shared Media Domains from the Embedded URLs}
\label{ssec:method_url}

URL sharing behavior is commonly used as a proxy to study media consumption patterns online~\cite{Bakshy2015ExposureTI}. To understand how Twitter users in different countries consume \covid news, we extract URLs embedded in the tweets posted by users whom we can estimate their geolocation and political leaning. We use the ``{\tt expanded\_url}'' field in the tweet object.

We find that a non-trivial number of URLs are shortened URLs. If the URLs are shortened by Twitter's own shortener (i.e., \textit{t.co}), the ``{\tt expanded\_url}'' field will contain the full URL address. If the URLs are shortened by other shortening services, we cross check the shortened URLs with a public shortener list.\footnote{\url{https://github.com/PeterDaveHello/url-shorteners}} We find \num{1061510} unique shortened URLs from 426 distinct shorteners. Among them, 331 shorteners (\num{314604} URLs) are platform-specific, meaning that the shortened URLs always redirect to 
a specific domain (e.g., \textit{cnb.cx} to \textit{CNBC}, \textit{wapo.st} to \textit{Washington Post}). The remaining 95 shorteners are general URL shortening services (e.g., \textit{bit.ly}, \textit{tinyurl.com}) and produce a total of \num{746906} shortened URLs. We programmatically send web requests to those shortened URLs and successfully resolve \num{532171} (71.3\%) full URL addresses. The overall success rate for resolving shortened URLs is 79.8\%, with failures due to HTTP timeouts, shorteners no longer functioning, and/or deprecated URLs.

We further extract the domain information from the shared URLs using a package called \texttt{tldextract}.\footnote{\url{https://github.com/john-kurkowski/tldextract}} There are two benefits of studying news consumption at the level of the media domain. First, if a URL has only been shared by a handful of users, aggregating it into the domain level would give us more users to analyze. Second, even from the same media group, news from different domains may target different audience bases. For example, one would expect that \textit{cnn} and \textit{cnnespanol} have very different audiences because the former focuses on English-speaking users while the latter focuses on Spanish-speaking users, though they both belong to the CNN media group. The finer data granularity is particularly desirable in cross-country analysis. In total, we obtain \num{4999798} unique URLs from \num{155302} distinct domains posted by \num{903699} Twitter users.

\section{Results}
\label{sec:result}

We start with showing the distribution of user political leaning in the eight countries (\Cref{ssec:result_ridge}). Next, we present an in-depth analysis of the bridging users who retweet from multiple countries (\Cref{ssec:result_bridging}). With the shared media domains from geolocated users with estimated political leaning, we derive a new media bias score by computing the average of audience leaning estimates, and we show this score is correlated with existing media bias scores (\Cref{ssec:result_biasscore}). Lastly, we profile \covid news consumption by generating the audience leaning distributions for a given country and for a given media domain (\Cref{ssec:result_consumption}).

\subsection{Distribution of User Political Leaning in Each Country}
\label{ssec:result_ridge}

\begin{figure*}[tbp]
  \centering
    \includegraphics[width=0.24\columnwidth]{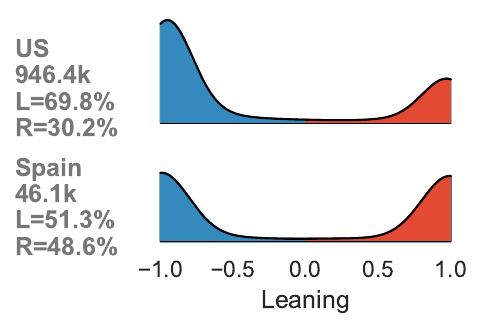}
    \includegraphics[width=0.24\columnwidth]{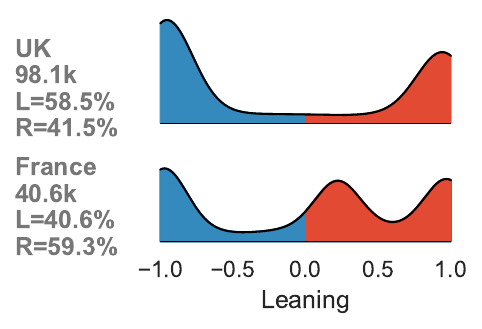} 
    \includegraphics[width=0.24\columnwidth]{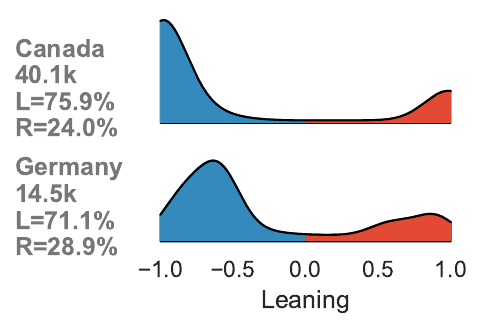} 
    \includegraphics[width=0.24\columnwidth]{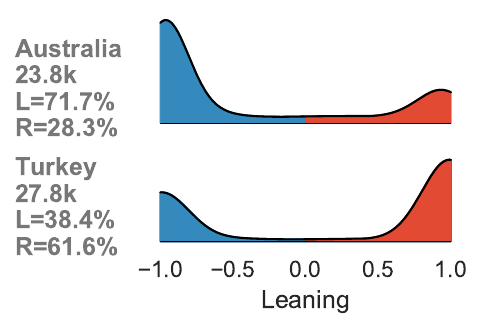} 
  \caption{The relative frequencies (density) of estimated user leaning in the eight countries. The numbers below a country name are the total number of geolocated users with estimated leaning scores, the fractions of left-leaning and right-leaning users found in the respective country from the \dataset dataset.} 
  \label{fig:ridge-plot}
\end{figure*}

\Cref{fig:ridge-plot} displays the density plots of the estimated user political leanings, one for each selected country. Users with scores in $[-1, 0)$ are considered leaning left within each country and colored in blue, while those in $(0, 1]$ are considered leaning right within each country and colored in red. Left-leaning users are the majority in most countries (ranging from 51.3\% to 75.9\%) except in France (\%L=40.6\%) and Turkey (\%L=38.4\%). For the US, the estimated fraction of left-leaning users is 69.8\%, and this is in line with a \covid era survey conducted by Pew Research Center finding that 69\% of prolific US Twitter users are pro-Democratic~\cite{pewtwitter}. Another overall observation is that the user distributions in seven out of eight countries have a bi-modal shape, meaning that only one mode (local maximum of a distribution) is in each of the left-leaning and right-leaning regimes. France is a notable exception with three modes in the user distribution.

\header{Interpretation and limits of interpretations.}
Twitter is known to operate more like broadcast media than a social network~\cite{kwak2010twitter}. We think that the literature on comparative media systems may help explain and conceptualize our observations. In the three models of Western media systems proposed by~\citet{hallin2004comparing}, the US, UK and Canada are classified as the \textit{liberal} model. We expect Australia to have a similar pattern due to the cultural resemblance and close alliance in the real world. The user distributions in these four countries indeed manifest high similarity (\Cref{fig:ridge-plot} top row) -- having a mode near the end of the left-leaning and right-leaning regimes. Yet there are also deviations from the comparative media system framework. For example, the media system in France is close to \textit{polarized pluralist}, with media integrated into party politics and a strong role of the state.

The three-mode user distribution we observe for France concurs with the commonly-used populist/mixed/non-populist groupings~\citep{pewfrance} although there may be a formal discussion on the three groups that we are not aware of. User distributions in our study appear highly polarized in almost all countries without a prominent center/middle group except for France.
The reason for this could be that social media inherently amplifies polarization~\cite{Bakshy2015ExposureTI}, and that
the original comparative media system framework pre-dates social media and only considers four dimensions: newspaper industry, political parallelism, media professionalized, and role of the state. This is confirmed by several recent literature, where \citet{powers2014internet} find that online media increases external pluralism in media content in the US. Overall, research has shown the positive role of social media in increasing people's political participation~\cite{mcclurg2003social}, but in the meantime, social media attracts considerable attention to increasing polarization of political views~\cite{hong2016political}, especially around \covid issues and policies~\cite{hegland2022partisan}.

Note that the user leanings are estimated on whom are connected to local politicians via retweets about \covid topics. While we do not claim that the studied user population is representative of each respective country or even its Twitter user base, in \Cref{sssec:biasscore_val}, we show that audience leaning scores estimated from this \dataset dataset have significant correlations with the media biases obtained via traditional survey methods in US, UK, Australia, France, Germany, and Spain~\cite{fletcher2020polarized}. We believe the geolocated users in \dataset dataset provide a novel 
and geographically diverse picture of political polarization around the world.

\subsection{An Analysis of the Bridging Users}
\label{ssec:result_bridging}

\begin{table*}[tbp]
    \centering
    \resizebox{\textwidth}{!}{
    \begin{tabular}{r|rrrrrrrrr} 
        \toprule
        & United States & United Kingdom & Canada & Australia & Spain & France & Germany & Turkey \\ 
        \midrule
        United States & \num{940154} & \num{2797} & \num{3924} & \num{2086} & \num{755} & \num{1076} & \num{1306} & \num{370} \\ 
        United Kingdom & \num{2150} & \num{95558} & \num{561} & \num{503} & \num{451} & \num{544} & \num{429} & \num{125} \\ 
        Canada & \num{1020} & \num{252} & \num{38954} & \num{147} & \num{61} & \num{109} & \num{91} & \num{26} \\ 
        Australia & \num{531} & \num{236} & \num{160} & \num{23186} & \num{40} & \num{76} & \num{82} & \num{33} \\ 
        Spain & \num{438} & \num{202} & \num{90} & \num{105} & \num{45561} & \num{170} & \num{148} & \num{10} \\ 
        France & \num{354} & \num{118} & \num{240} & \num{46} & \num{128} & \num{40026} & \num{139} & \num{31} \\ 
        Germany & \num{285} & \num{106} & \num{68} & \num{44} & \num{73} & \num{90} & \num{14157} & \num{67} \\ 
        Turkey & \num{173} & \num{68} & \num{44} & \num{55} & \num{0} & \num{115} & \num{154} & \num{27504} \\
        \bottomrule 
    \end{tabular}
    }
    \caption{The number of geolocated users from the \{row\} country who have retweeted at least once geolocated users from the \{column\} country. The off-diagonal entries are the sets of bridging users between the \{row\} country and \{column\} country.}
    \label{tab:bridge}
\end{table*}

One key challenge in comparative research is the systematic variation of data collected from different places and/or different periods. Putting it into the context of cross-country political polarization, it is possible that left-leaning users in one country are more \textit{right} than right-leaning users in another country. This misalignment would invalidate observations at face value. To make fair comparisons, a common axis is needed -- a process we call ``calibration''. Traditionally, researchers rely on a set of users, whom they can collect answers along two dimensions and learn a mapping function between these two dimensions. They can then apply the mapping function to calibrate answers from one dimension to the other dimension or vice versa. For instance, \citet{lo2014common} surveyed respondents about perceptual placements of multiple parties in Europe and calibrated the results to construct a common left-right scale to correct party positions across countries.
Recall that the political leaning scores we estimate in~\Cref{ssec:method_leaning} quantify how frequently a user retweets information from the left and right-leaning groups. The scores do not measure the degree of leaning left or right. We therefore cannot answer questions such as \textit{are US left-leaning users more right than UK right-leaning users}. Instead, we should ask \textit{do US left-leaning users retweet more from UK right-leaning users than from UK left-leaning users}. If the answer is yes, we should design a calibration procedure to align US left-leaning users with UK right-leaning users.

We identify a set of users who have interactions between multiple countries. We define \textit{bridging users} as the geolocated users from one country and having retweeted geolocated users from another country. \Cref{tab:bridge} summarizes the numbers of bridging users for every country pair. First, the diagonal entries show that most users' retweeting behaviors are limited to other users in the same country.
Second, we focus on the bridging users between US and non-US countries due to the low number of bridging users between any pair of two non-US countries in \dataset. For instance, we find \num{2797} US users retweeting UK users and \num{2150} UK users retweeting US users; hence a total of 4947 bridging users between US and UK. We generate two versions of political leaning scores for the bridging users. Specifically, for bridging users between the US and a country $C$, we first add them to the within-country retweet networks from the US and from $C$. We then repeat the label spreading step to estimate their respective leaning scores by using the networks from the US and from $C$. We hence obtain two predictions for each bridging user. 

\begin{figure*}[tbp]
  \centering
  \captionsetup[subfigure]{labelformat=empty}
  \subcaptionbox{}[.24\columnwidth][c]{%
    \includegraphics[width=.24\columnwidth]{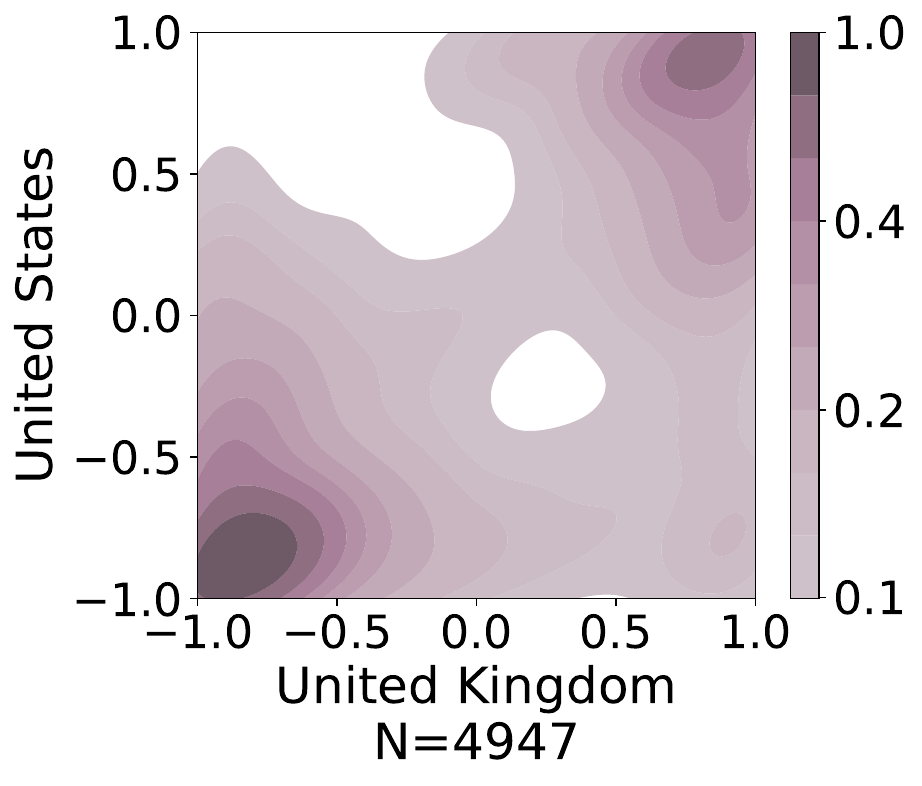}}
  \subcaptionbox{}[.24\columnwidth][c]{%
    \includegraphics[width=.24\columnwidth]{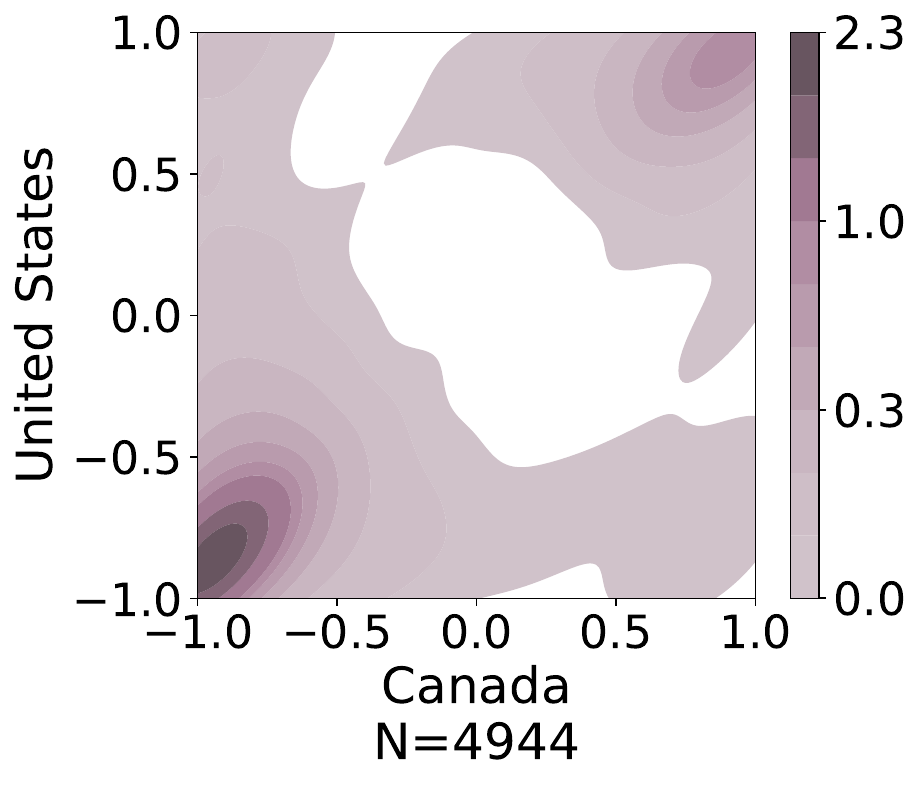}}
  \subcaptionbox{}[.24\columnwidth][c]{%
    \includegraphics[width=.24\columnwidth]{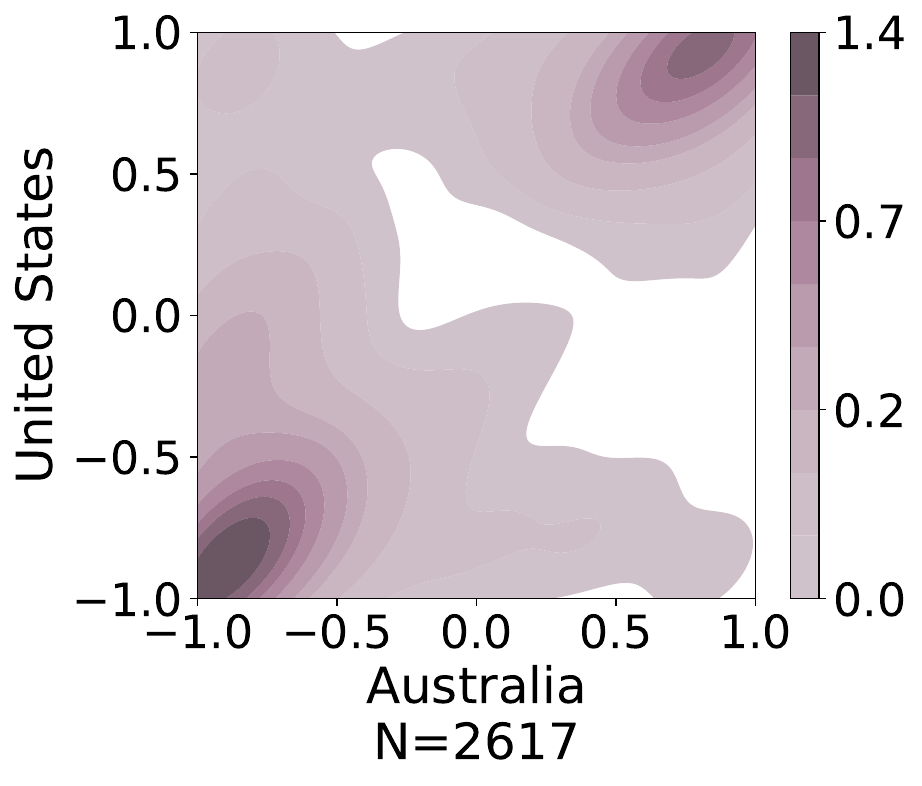}} \\
  \subcaptionbox{}[.24\columnwidth][c]{%
    \includegraphics[width=.24\columnwidth]{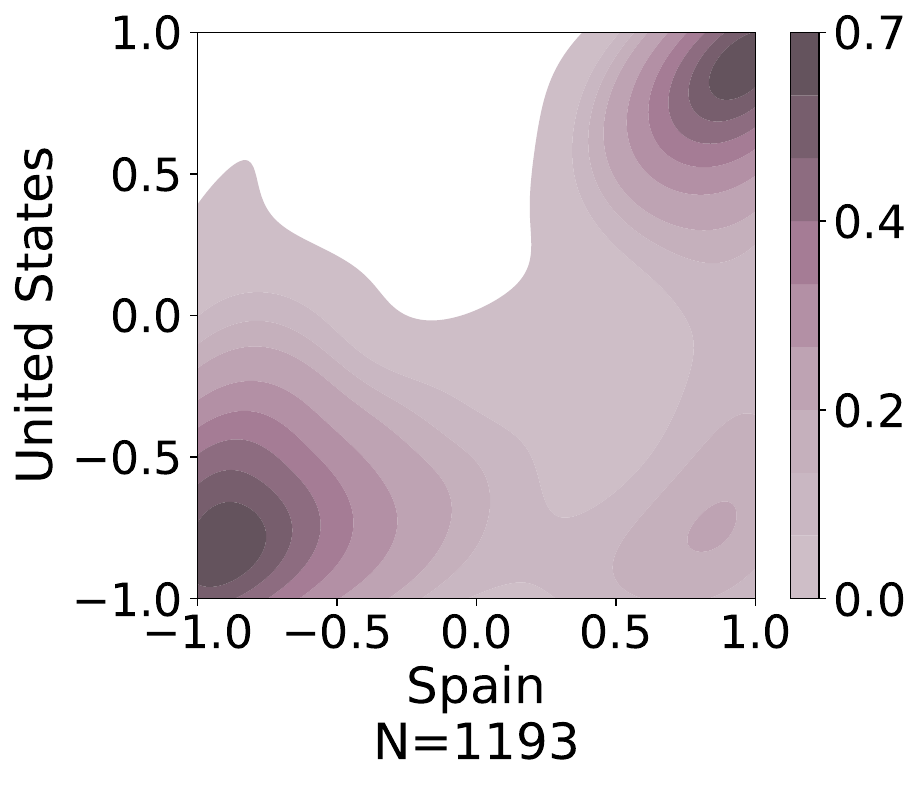}}
  \subcaptionbox{}[.24\columnwidth][c]{%
    \includegraphics[width=.24\columnwidth]{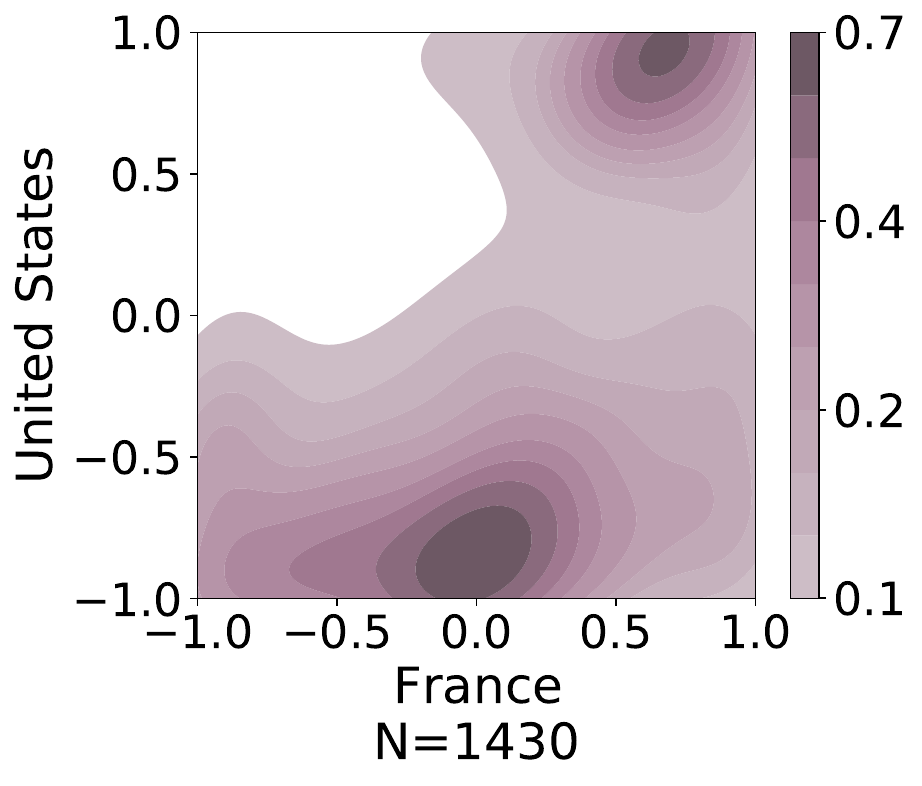}}
  \subcaptionbox{}[.24\columnwidth][c]{%
    \includegraphics[width=.24\columnwidth]{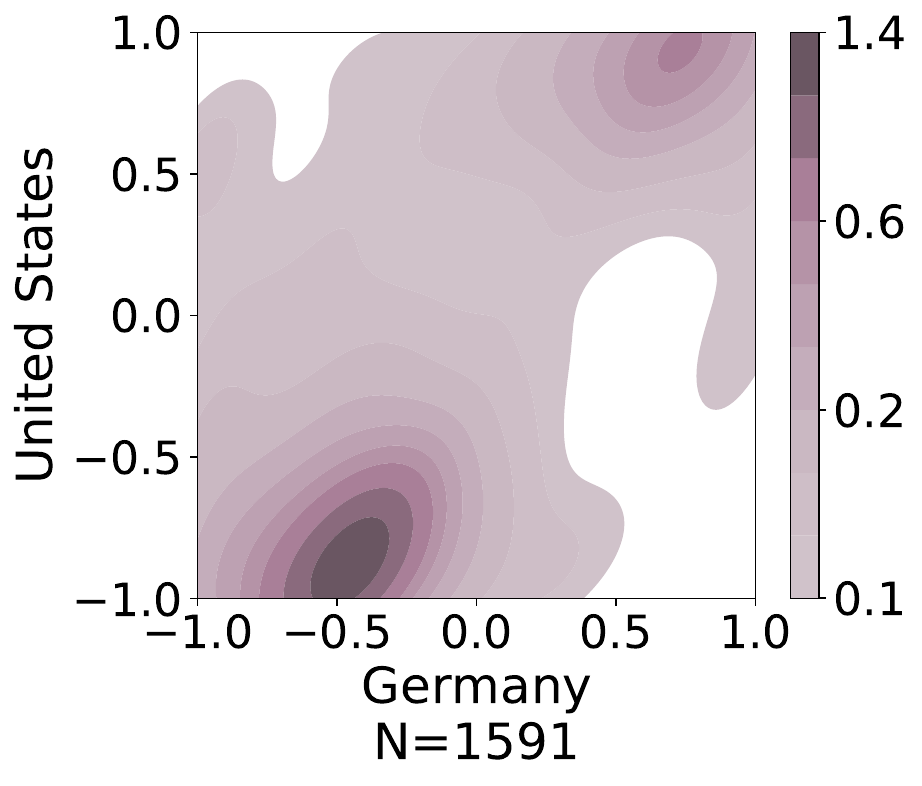}}
  \subcaptionbox{}[.24\columnwidth][c]{%
    \includegraphics[width=.24\columnwidth]{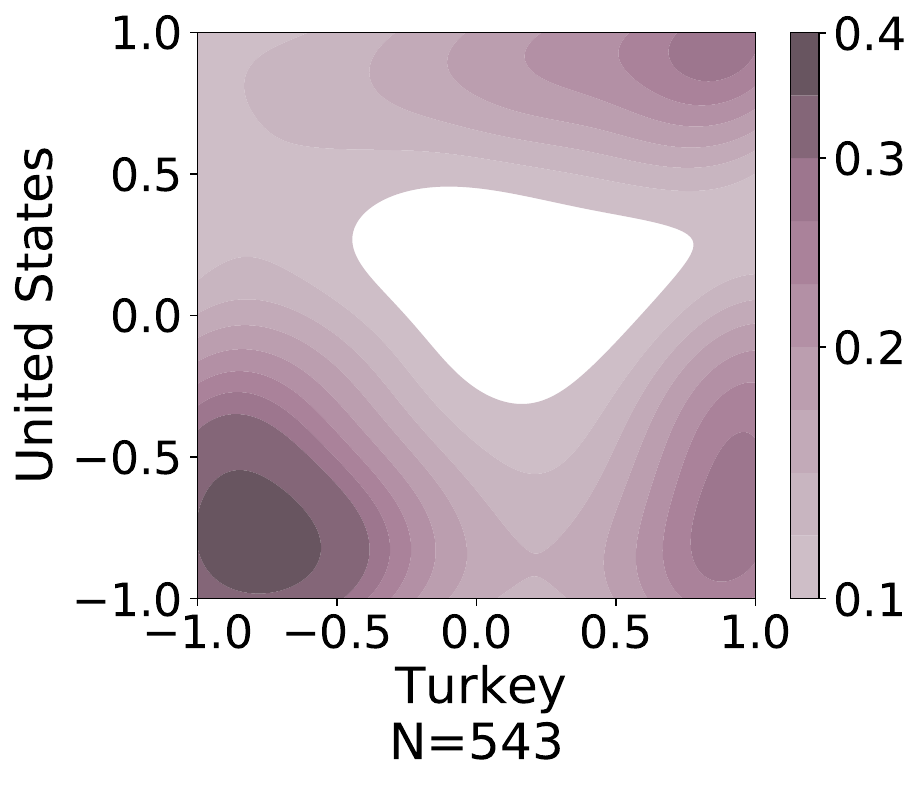}}
   \caption{2D density plot of two versions of predicted political leaning scores for bridging users.
   x-axis: estimated by using the user-user retweet network in \{x\_label\} country; y-axis: by using the retweet network in the US. $N$ indicates the size of bridging users in each country pair. Except for France and Turkey, all other countries have the highest densities in the bottom left and top right regimes.}
   \label{fig:bridging-users}
\end{figure*}

\header{Why do we NOT calibrate the predicted scores across countries?}
\Cref{fig:bridging-users} shows the 2D density plot of the two versions of predicted scores in each US vs. non-US country pair.
We find that densities are higher in the bottom left and top right area in almost all pairs, indicating consistent retweeting behaviors across countries -- if a group of users mostly retweet one political side in a country, they would not switch to mostly retweeting the opposite political side in another country. This observation warrants that no special calibration is needed to adjust the political leaning scores estimated from different within-country retweet networks. Even if there exists a latent common axis that we can project all predicted scores onto, we are unlikely to overturn the estimated left and right-leaning labels.
Two exceptions in \Cref{fig:bridging-users} are France and Turkey. For the France-US pair, many bridging users interact with a politically diverse set of French users while mostly interacting with left-leaning US users. For the Turkey-US pair, although the highest density still occurs in the bottom left area, the bottom right area also has a moderate density, suggesting a modest number of users who mainly interact with right-leaning Turkish users and left-leaning US users. This may relate to two prominent groups of right-leaning French users and a large number of right-leaning Turkish users as presented in~\Cref{fig:ridge-plot}. In order to understand the nuances there, one has to investigate the tweets they posted and the users they retweeted. We leave this case study as future work.

We also experimented with fitting a linear regression function for each country pair similar to \citet{lo2014common}. We did not report the calibrated results for two reasons: (1) the 2D density plot suggests the likelihood of a non-linear relation; hence the linear fitting is not ideal; (2) the density in the middle area is low; hence we would have a high level of uncertainty (i.e., large confidence interval) for users around that area. 

\subsection{A New Media Bias Score by Average Audience Leaning}
\label{ssec:result_biasscore}

\subsubsection{Computing audience leaning distribution for each domain}
\label{sssec:biasscore_comp}

In this work, each Twitter user can be represented as a tuple $(u, l_u, p_u, D_u)$, where $u$ is the user id, $l_u$ is the extracted user location (\Cref{ssec:method_geo}), $p_u$ is the estimated political leaning score (\Cref{ssec:method_leaning}), and $D_u = \{d_{u1}, d_{u2} \ldots\}$ is the set of media domains that user $u$ has shared (\Cref{ssec:method_url}).

\begin{figure*}[tbp]
  \centering
  \includegraphics[width=.98\columnwidth]{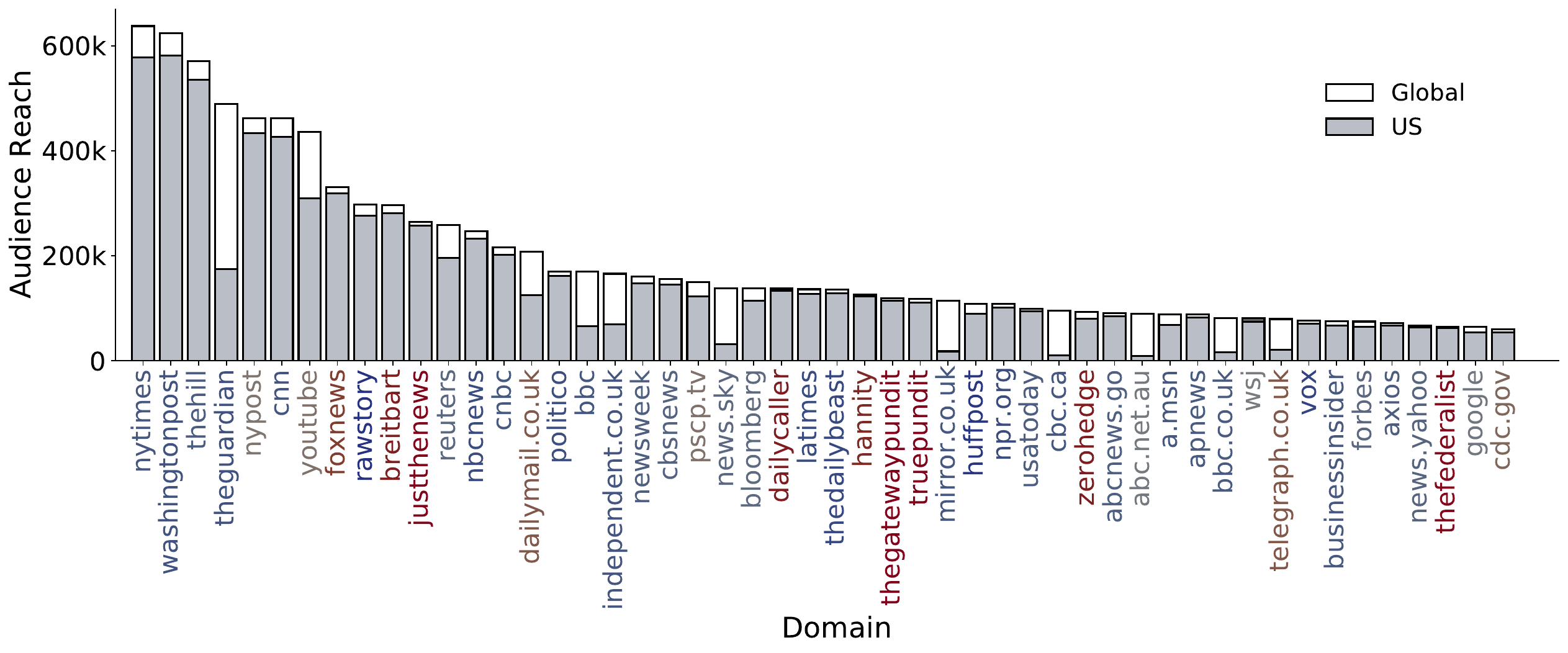}  
  \caption{Top 50 domains ranked by their global audience reach. The shaded area indicates the size of the US audience. The domain names are colored by their average audience leaning scores in the US. Blue (red) color indicates more left-leaning (right-leaning) audience.}
  \label{fig:domain-ranking}
\end{figure*}

We define the global {\it audience reach} $\kappa(d)$ of a media domain $d$ as the number of all unique Twitter users sharing URLs from this domain $\kappa(d)=\left| \{u | d\in D_u\} \right|$. Similarly, the audience reach of domain $d$ in location $l$ is $\kappa(d, l)=\left| \{u | d\in D_u, l_u=l\} \right|$, where $\left| \cdot \right|$ denotes set cardinality. \Cref{fig:domain-ranking} displays the top 50 domains according to their global audience reach.\footnote{Despite using the word ``media'', we notice that some platforms also appear in top positions (e.g., \textit{youtube}, \textit{pscp.tv}, \textit{google}). They are content hosting services rather than producing news content by themselves. We opt to keep those platform domains since many media outlets also function as a platform for user-generated content (e.g., \textit{New York Times Opinion}), and drawing a precise line is difficult.} We observe not only mainstream media (e.g., \textit{cnn}, \textit{foxnews}), but also alternative, extreme media (e.g., \textit{breitbart}, \textit{thegatewaypundit}). The shaded area indicates the size of the US audience. We find many non-US-based media (e.g., \textit{theguardian} from the UK, \textit{cbc.ca} from Canada). The top domains span the full range of the left-right political spectrum (e.g., from left-leaning media \textit{cnn} and \textit{bbc}, to right-leaning media \textit{breitbart} and \textit{independent.co.uk}). The domain names are colored by their average audience leaning scores (detail follows), where blue (red) indicates a more homogeneous left-leaning (right-leaning) audience base.

We compute the {\it average audience leaning} $\bar p(d)$ of each domain $d$ by averaging over all users who have shared $d$ at least once. Similarly, we can compute the average leaning of domain $d$ shared by users in location $l$, denoted as $\bar p(d, l)$. In the subsequent use of the average audience leaning score, we omit the notions of domains $d$ and location $l$ when they are clear from context. 

$$\bar p(d) = \frac{1}{\kappa(d)}\sum_{u | d \in D_u} p_u; ~\hspace{1cm}~\bar p(d, l) = \frac{1}{\kappa(d, l)}\sum_{u | d \in D_u, l_u=l} p_u $$

User-based aggregation was commonly used in recent work, but only to compute the mean. \citet{Bakshy2015ExposureTI} and \citet{robertson2018auditing} both described a domain by averaging over the leaning scores of users who had shared it. 
Information extracted from the \dataset dataset offers a rich set of statistics for each domain. The distribution of user leanings for domain $d$ in location $l$ can be represented as mean, median, and density plots (See plotting styles of \Cref{fig:ridge-plot}, \ref{fig:top15-domain} and \ref{fig:domain-across-countries}).

\begin{figure*}[tbp]
  \centering
  \subcaptionbox{\centering Pew Research Center~\cite{pewscores}, n=24 (86\%), $r=0.94^{***}$ \label{subfig:pew} }[.32\columnwidth][c]{\includegraphics[width=.32\columnwidth]{ 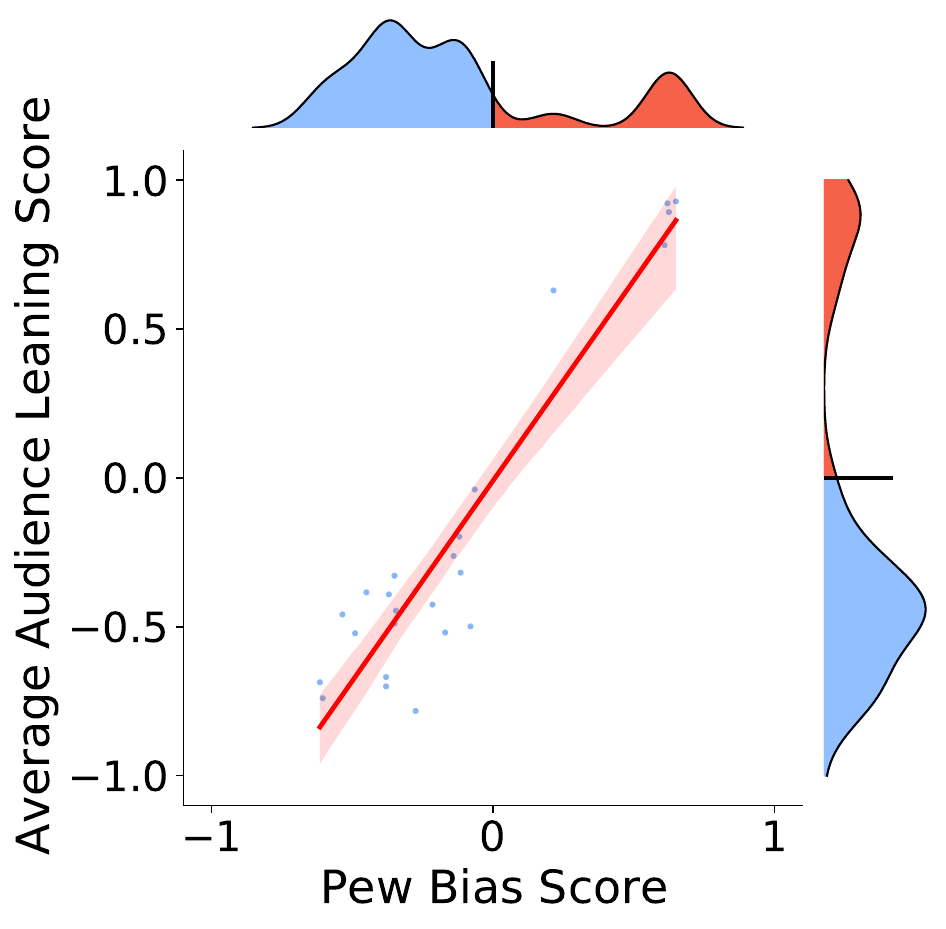 } }
  \subcaptionbox{\centering \citet{Bakshy2015ExposureTI}, n=281 (65\%), $r=0.88^{***}$ \label{subfig:bakshy} }[.32\columnwidth][c]{\includegraphics[width=.32\columnwidth]{ 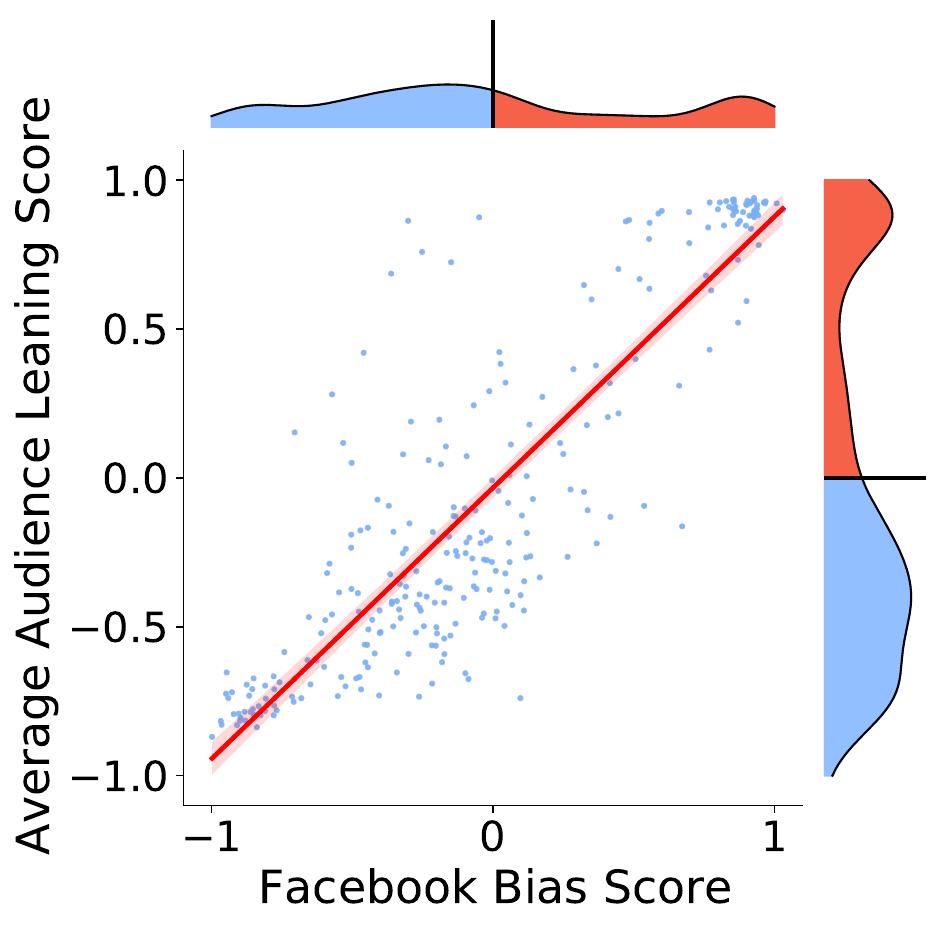 } }
  \subcaptionbox{\centering \citet{Budak2016FairAB}, n=14 (93\%), $r=0.83^{***}$ \label{subfig:budak} }[.32\columnwidth][c]{\includegraphics[width=.32\columnwidth]{ 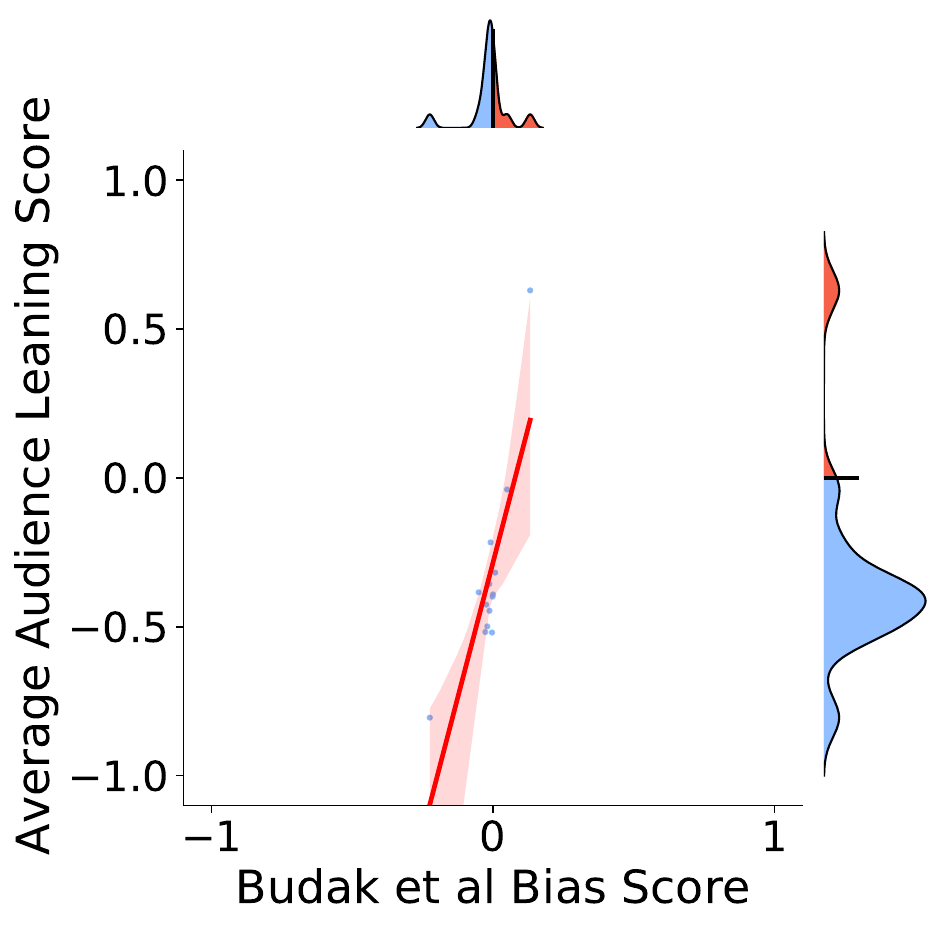 } } \\
  \subcaptionbox{\centering \citet{allsides}, n=237 (81\%), $\rho=0.75^{***}$ \label{subfig:allsides} }[.32\columnwidth][c]{\includegraphics[width=.32\columnwidth]{ 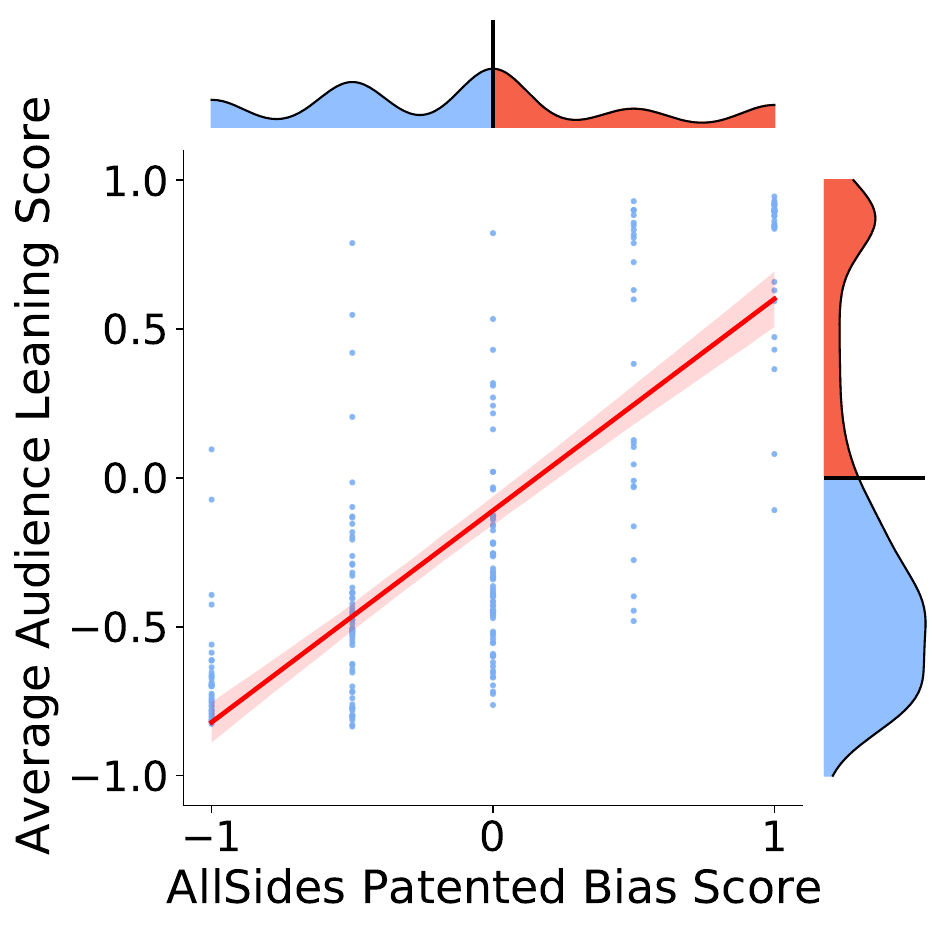 } }
  \subcaptionbox{\centering \citet{robertson2018auditing}, n=3132 (16\%),  $r=0.63^{***}$ \label{subfig:rob} }[.32\columnwidth][c]{\includegraphics[width=.32\columnwidth]{ 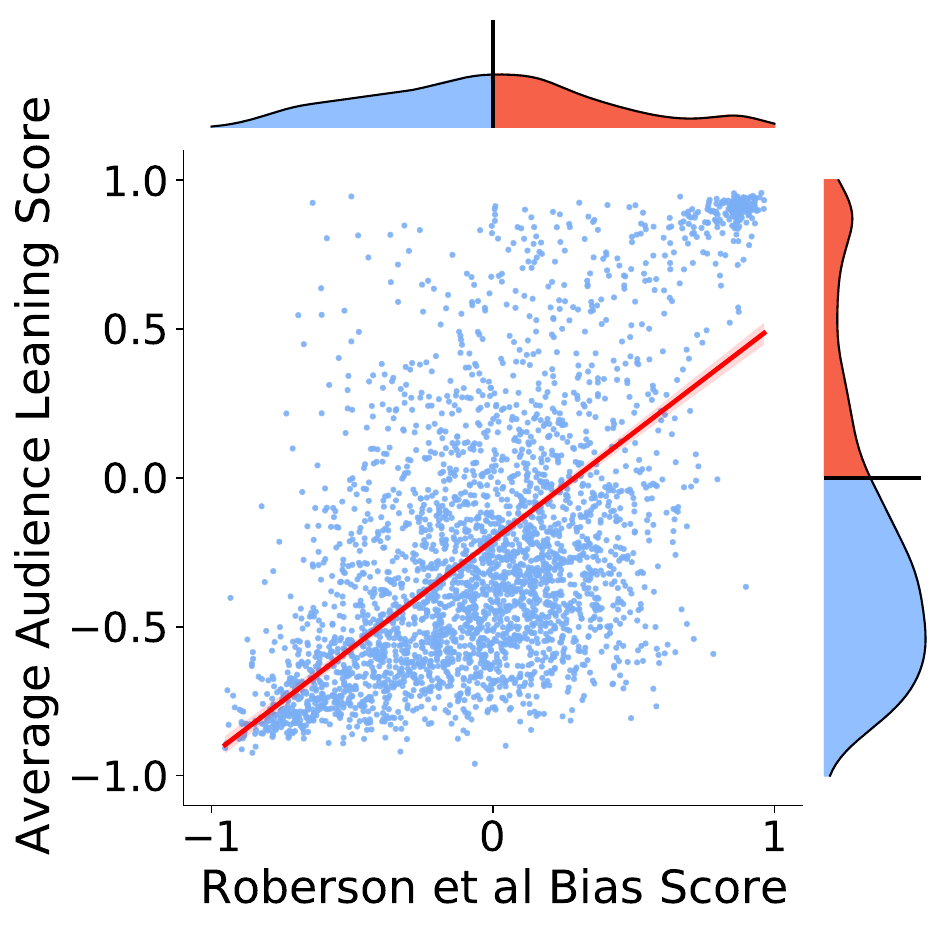 } }
  \subcaptionbox{\centering \citet{robertson2018auditing}, n=262 (68\%), $\rho=0.46^{***}$ \label{subfig:rob-mturk} }[.32\columnwidth][c]{\includegraphics[width=.32\columnwidth]{ 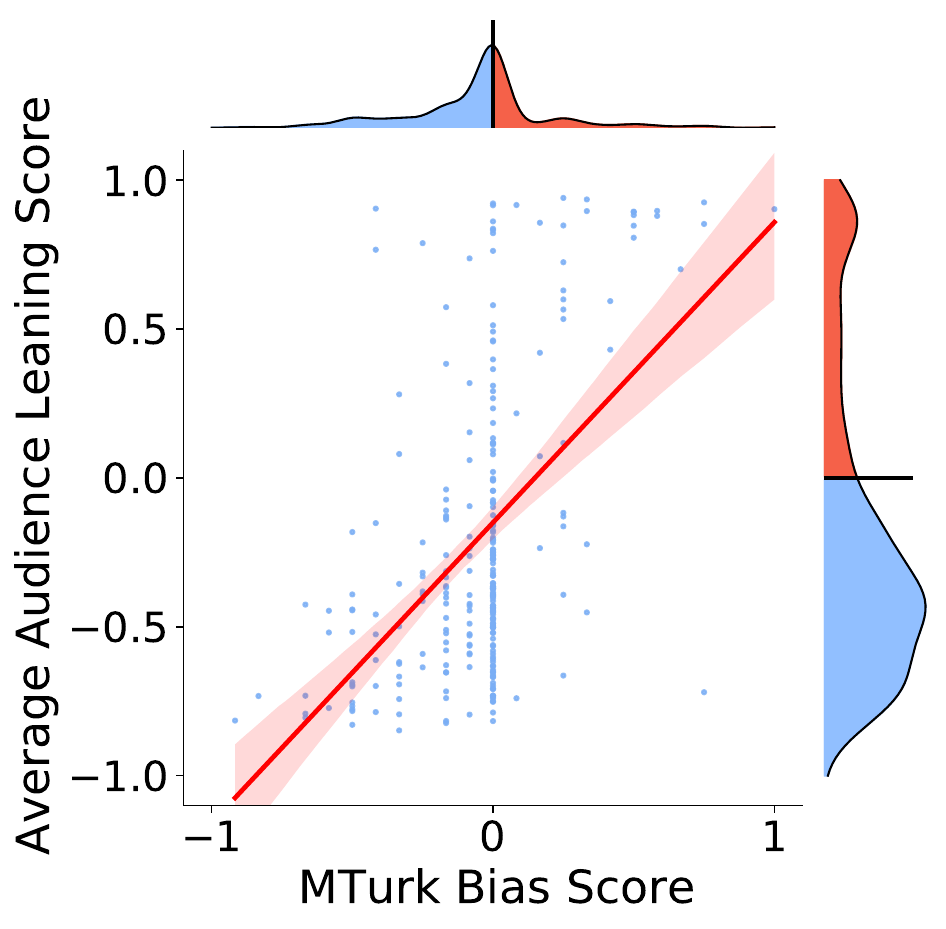 } }
  \caption{Correlation between average audience leaning scores from our data (y-axis) vs. media bias scores from prior literature (x-axis). The subfigure title lists the reference, number of overlapped domains, coverage, and correlation coefficient (Pearson's $r$ for numerical labels and Spearman's $\rho$ for ordinal labels). $^{***}p < 0.001$. Side densities represent the distribution of domain leaning scores from either source. }
  \label{fig:us-other-scores}
\end{figure*}

\begin{figure*}[tbp]
  \centering
  \subcaptionbox{\centering n=24 (100\%), $r=0.88^{***}$ \label{subfig:fletcher-us} }[.32\columnwidth][c]{\includegraphics[width=.32\columnwidth]{ 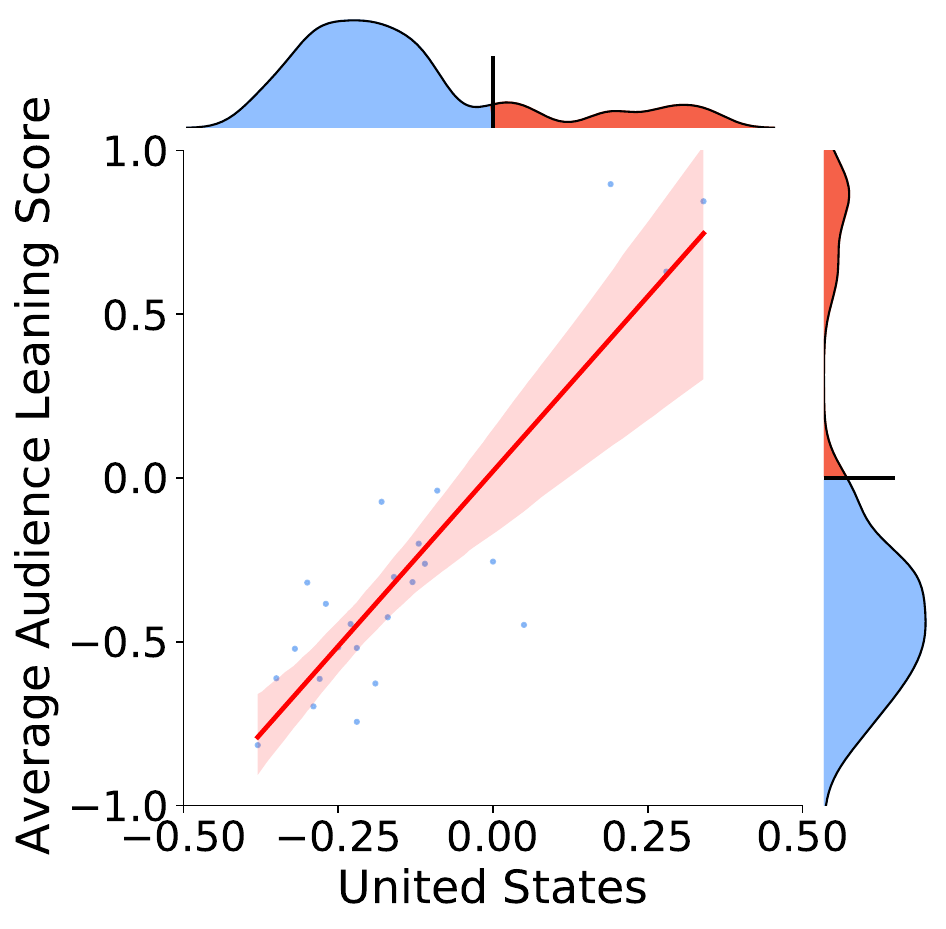 } }
  \subcaptionbox{\centering n=19 (100\%), $r=0.80^{***}$ \label{subfig:fletcher-uk} }[.32\columnwidth][c]{\includegraphics[width=.32\columnwidth]{ 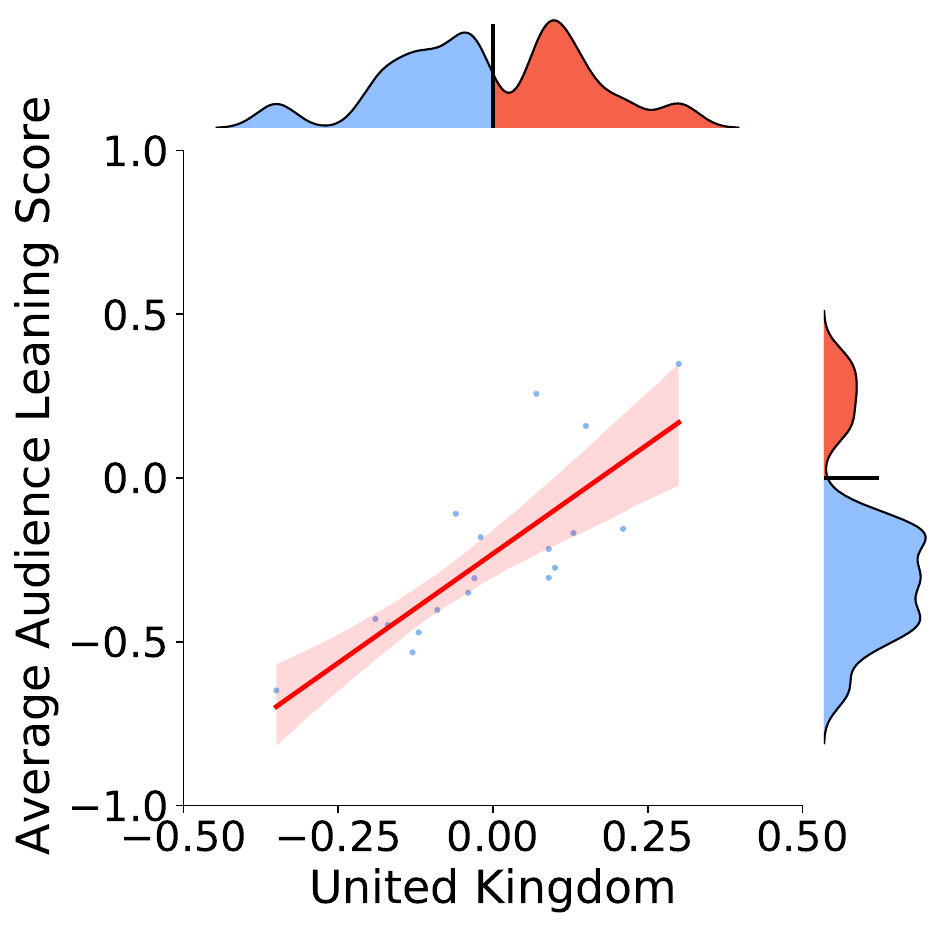 } }
  \subcaptionbox{\centering n=21 (100\%), $r=0.81^{***}$ \label{subfig:fletcher-au} }[.32\columnwidth][c]{\includegraphics[width=.32\columnwidth]{ 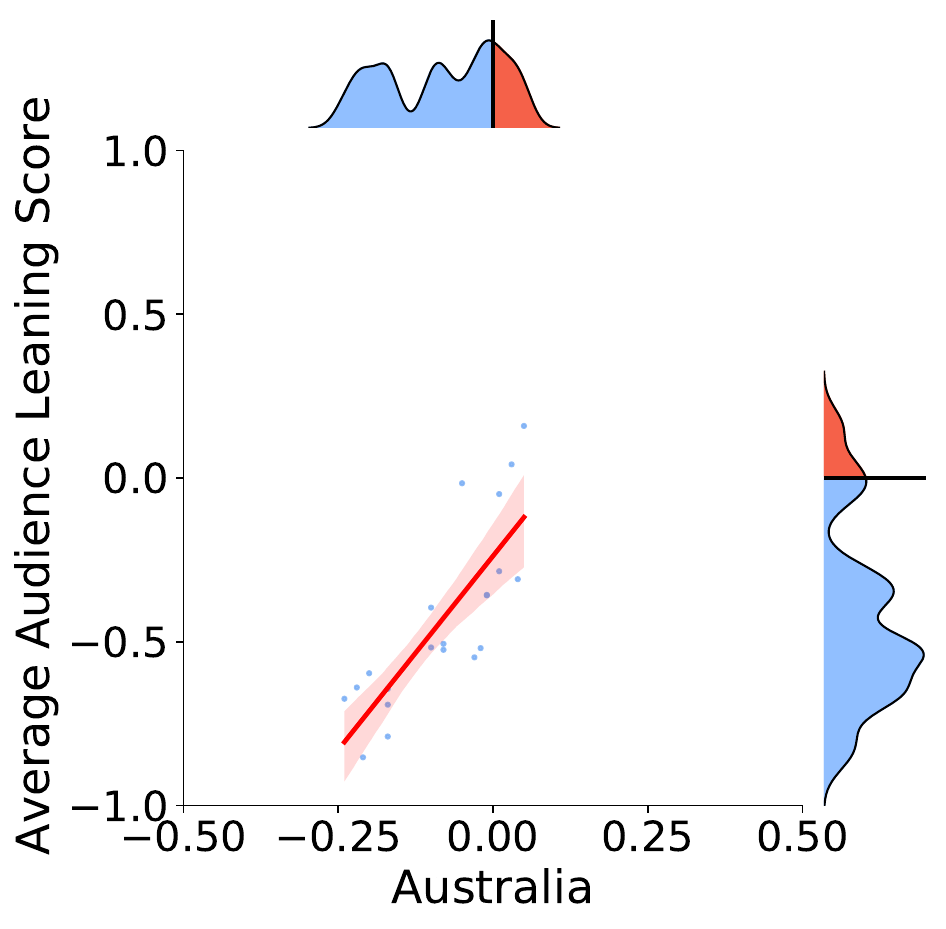 } } \\
  \subcaptionbox{\centering n=28 (100\%), $r=0.75^{***}$ \label{subfig:fletcher-es} }[.32\columnwidth][c]{\includegraphics[width=.32\columnwidth]{ 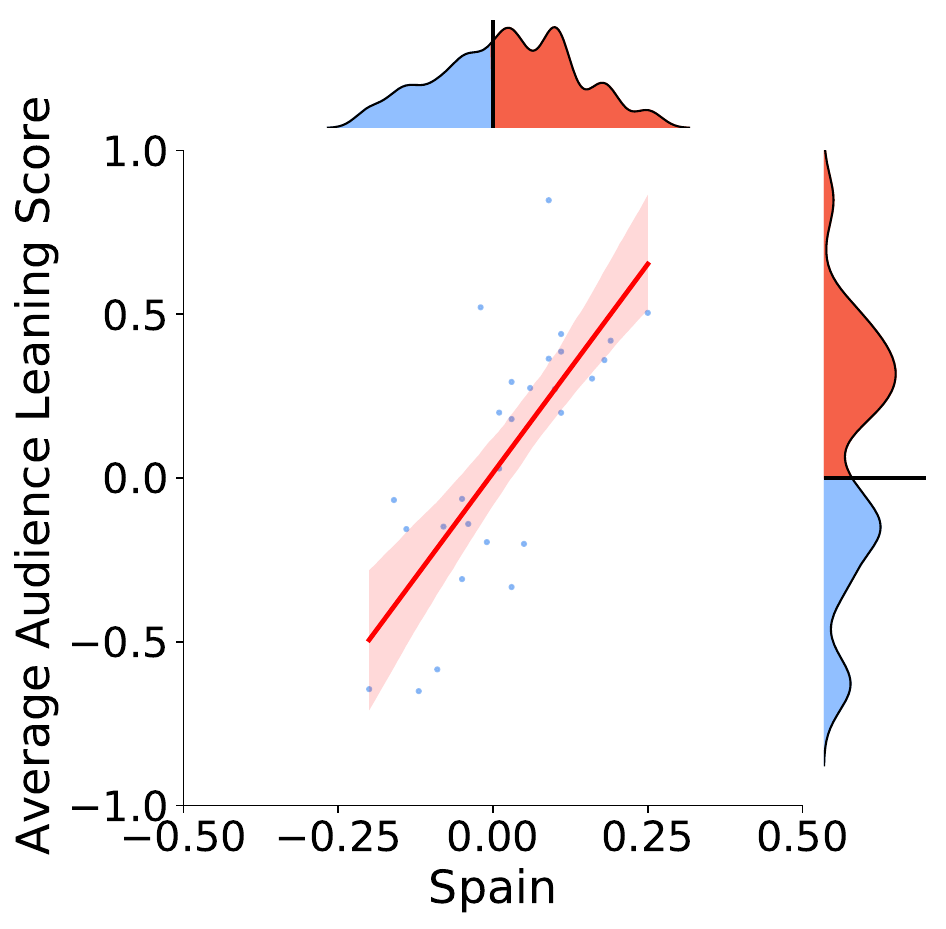 } }
  \subcaptionbox{\centering n=26 (100\%), $r=0.72^{***}$ \label{subfig:fletcher-fr} }[.32\columnwidth][c]{\includegraphics[width=.32\columnwidth]{ 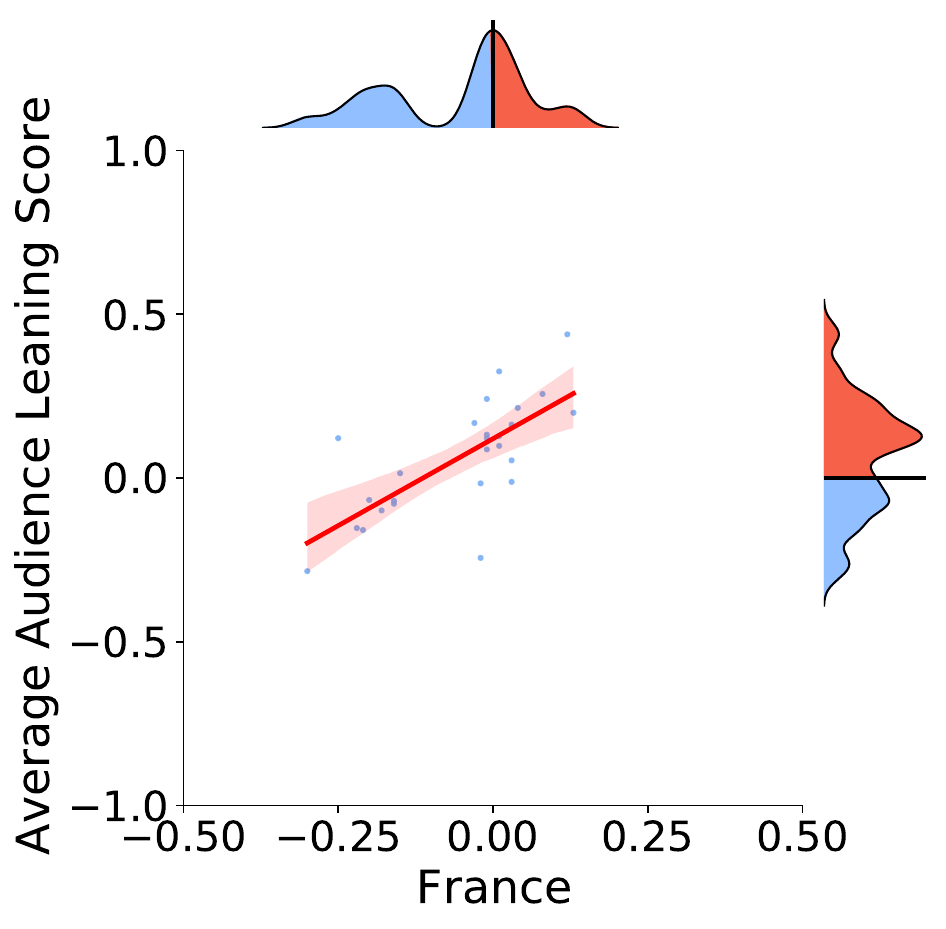 } }
  \subcaptionbox{\centering n=19 (95\%), $r=0.50^{*}$ \label{subfig:fletcher-de} }[.32\columnwidth][c]{\includegraphics[width=.32\columnwidth]{ 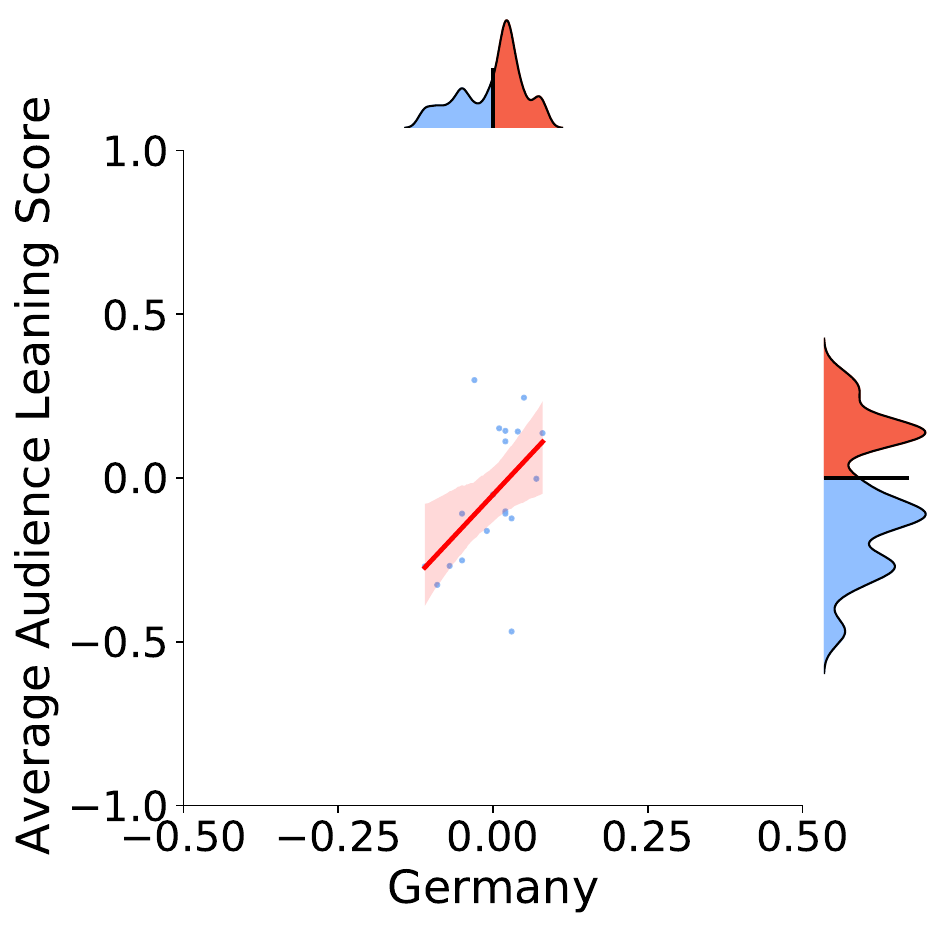 } }
  \caption{Correlation between our average audience leaning score (y-axis) vs. those from \citet{fletcher2020polarized} (x-axis). The subfigure title lists the number of overlapped domains, coverage, and Pearson's $r$ correlation test. $^{***}p < 0.001, ^{*}p < 0.05$. For domains shared in multiple countries, we calculate the average leaning scores from audience in each country independently.}
  \label{fig:fletcher}
\end{figure*}

\subsubsection{Validating average audience leanings}
\label{sssec:biasscore_val}
As a validation, we compare the average audience leaning to two kinds of publicly available media bias reporting. 


\header{Comparing to existing media bias ratings.}
We compare the average audience leaning scores to six other estimates from recent literature~\cite{Bakshy2015ExposureTI, Budak2016FairAB, pewscores, robertson2018auditing,allsides}. Since these studies focus on the US users, we compare them to the average audience leaning for US users $\bar p(d, l=\text{`US'})$. To mitigate noise, we only include domains with a US audience reach of at least 50. This yields \num{7924} media domains. We take the intersecting domains from our study and prior literature. Some work provides a numerical label (from -1 to 1), while some work provides an ordinal label (from extreme left to center to extreme right).
We compute Pearson's correlation coefficient $r$ for numerical labels and Spearman's rank correlation coefficient $\rho$ for ordinal labels. The higher the coefficient is, the stronger the correlation between the two data sources. \Cref{fig:us-other-scores} summarizes the results with scatter plots of the scores and correlation coefficients. 

Pew Research's audience profile scores~\cite{pewscores} were collected from \num{2901} web respondents that were representative of the US Internet users in 2014. We reconstructed these scores from the interactive webpage~\cite{pewscores2}. The remaining five sets of media bias scores are obtained from \cite{robertson2018auditing,Bakshy2015ExposureTI,Budak2016FairAB,allsides}.  
\Cref{subfig:pew,subfig:bakshy,subfig:budak,subfig:allsides} show high correlations between $\bar p$ and survey-based~\cite{pewscores}, sharing-based~\cite{Bakshy2015ExposureTI}, crowdsourcing-based~\cite{Budak2016FairAB}, and expert-based~\cite{allsides} media bias scores. Notably, the correlation of $\bar p$ with the Pew scores and AllSides scores are much higher ($r=0.94^{***}$ and $\rho=0.75^{***}$ respectively) than the same comparisons conducted in~\cite{robertson2018auditing} ($r=0.78^{***}$ and $\rho=0.64^{***}$ respectively). The lowest correlation is observed with the MTurk scoring ($\rho=0.46^{***}$ in \Cref{subfig:rob-mturk}), but it is consistent with~\citet{robertson2018auditing}'s own observation ($r=0.50^{***}$).

\header{Comparing to international media surveys.}
A key contribution of our study is the cross-country analysis. To this end, we compare $\bar p$ with survey results conducted by \citet{fletcher2020polarized}, in which respondents are a stratified sample from 12 different countries, and were asked about their political leaning on a seven-point Likert scale (later coded into a scale from $-0.5$ to $0.5$). Respondents were also asked about the news outlets they had read online and offline in the past weeks, from a candidate list of 30 popular outlets that varied from country to country. 
\citet{fletcher2020polarized}'s political leaning score for each news outlet is the mean of the self-identified leaning scores of its audience. We find a strong correlation between the media bias scores from our estimation and that from \cite{fletcher2020polarized} in the US, UK and Australia ($r \geq 0.8^{***}$, \Cref{subfig:fletcher-us,subfig:fletcher-uk,subfig:fletcher-au}). We also notice a moderate correlation in Spain and France ($r \geq 0.7^{***}$, \Cref{subfig:fletcher-fr,subfig:fletcher-es}). The correlation with Germany is lower ($r=0.50^{*}$, \Cref{subfig:fletcher-de}). This is likely due to that survey outcomes for German media in~\cite{fletcher2020polarized} are mostly around the center.

Through two validation tasks, we find strong correlations between our computed average audience leaning score $\bar p$ and other estimates of US media domain biases and international surveys. This provides us great confidence in using the new scores to produce novel observations about the media consumption behavior internationally, though we caution that the reliability of $\bar p$ in countries other than those validated above may require further scrutiny.

\subsection{Profiling Cross-Country \covid News Consumption}
\label{ssec:result_consumption}

\begin{figure*}[tbp]
  \centering
   \includegraphics[width=.99\columnwidth]{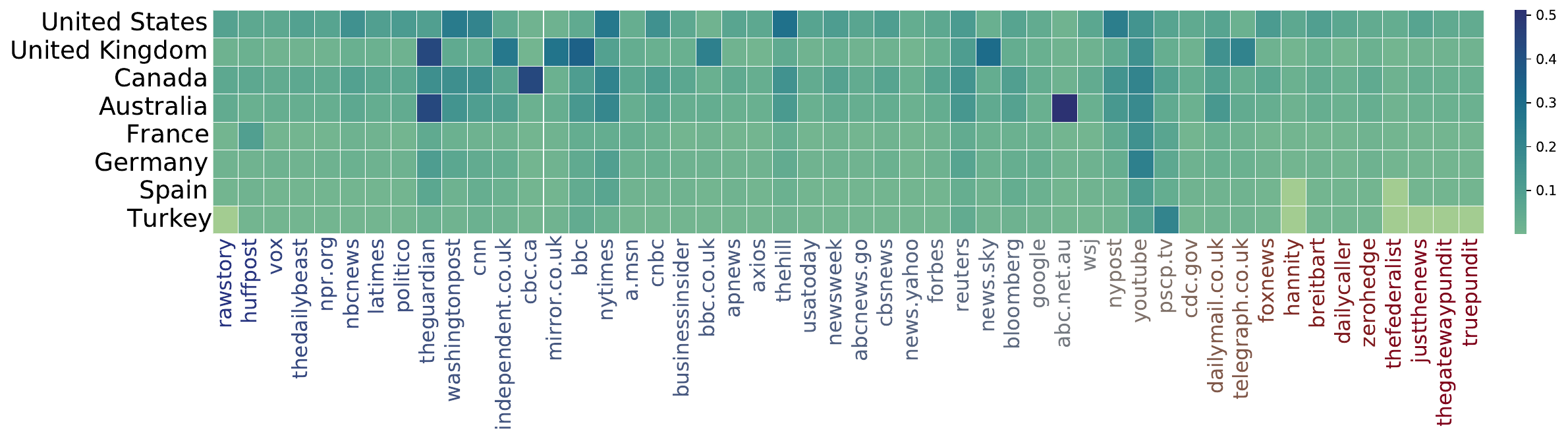}
   \caption{Heatmap of within-country audience reach of the overall top 50 domains in the eight selected countries. x-axis: domain names ranked by their average audience leaning in the US. y-axis: country names. Cell color represents the fraction of audience reach domain $d$ in country $l$ ($\kappa(d, l)$), normalized by all users in that country.}
  \label{fig:heatmap}
\end{figure*}

\Cref{fig:heatmap} provides an overview of the audience reach of the top 50 media domains (ordered by average audience leaning in the US) across the eight selected countries. These background statistics confirm the popularity of mainstream media in the four English-speaking countries, such as the {\it guardian, bbc, news.sky} for the UK, {\it cbc.ca} for Canada, and {\it guardian} and {\it abc.net.au} for Australia. Note that the top 50 domains are dominated by those in English, and that media consumption in the four non-English-speaking countries does not have a prominent peak -- necessitating country-specific profiles that examine non-English domains that are important for each country. This motivates the next set of profile plots that visualize audience distributions in a way comparable across media or across countries.

\begin{figure*}[!htbp]
    \centering
    \captionsetup[subfigure]{labelformat=empty}
    \subcaptionbox{}[.49\columnwidth][c]{%
    \includegraphics[width=.49\columnwidth]{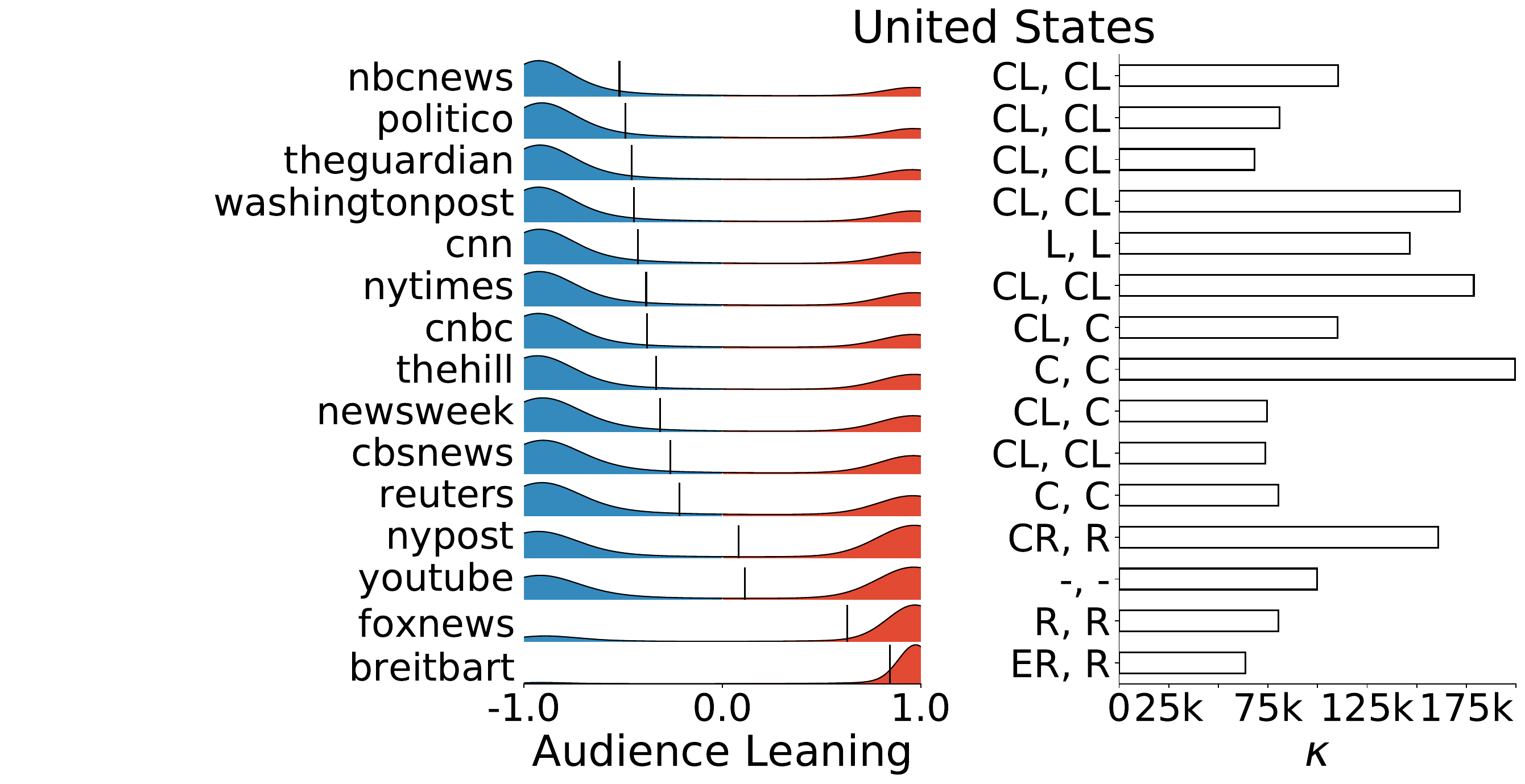}}
    \subcaptionbox{}[.49\columnwidth][c]{%
    \includegraphics[width=.49\columnwidth]{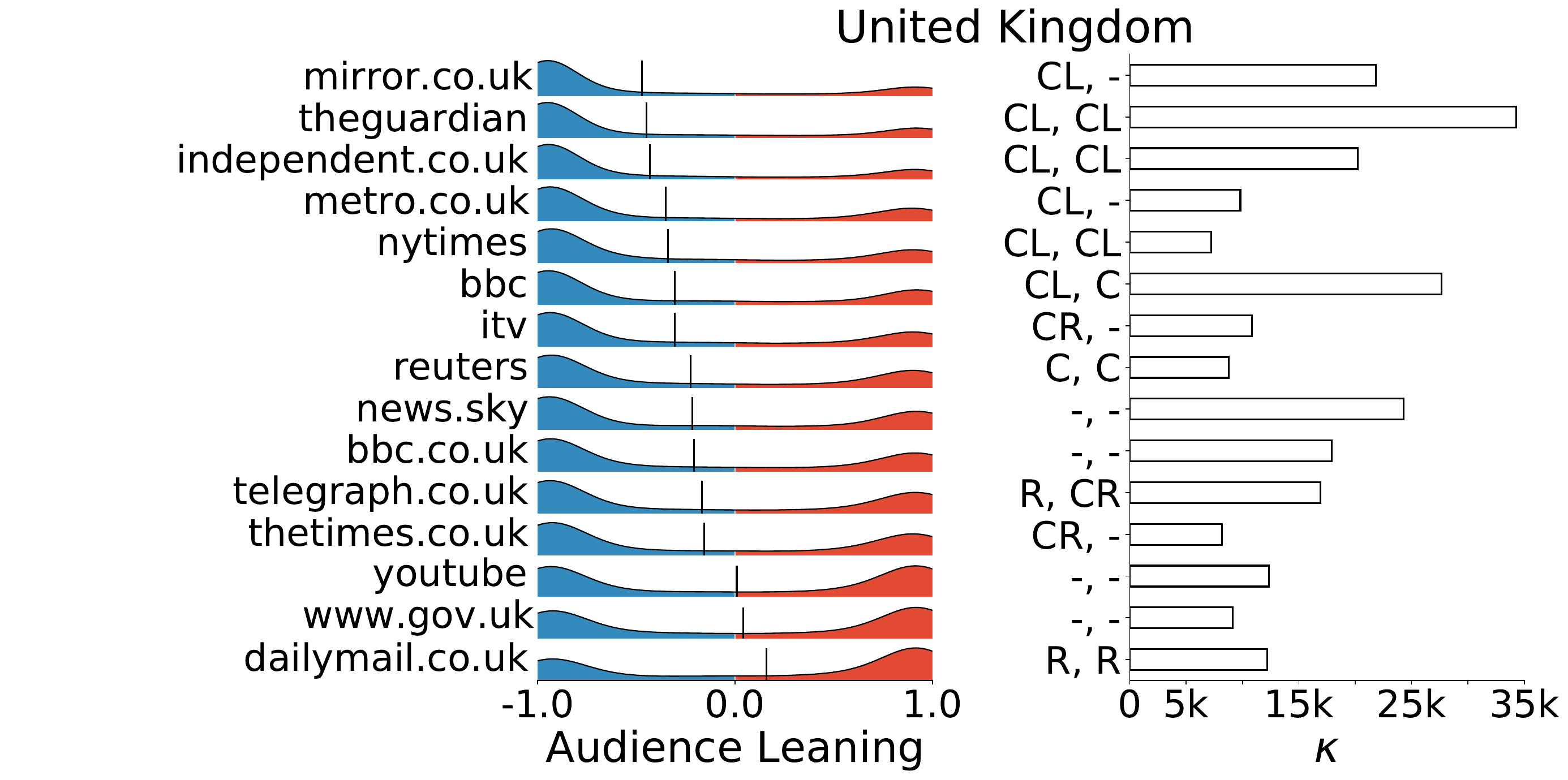}}
    \subcaptionbox{}[.49\columnwidth][c]{%
    \includegraphics[width=.49\columnwidth]{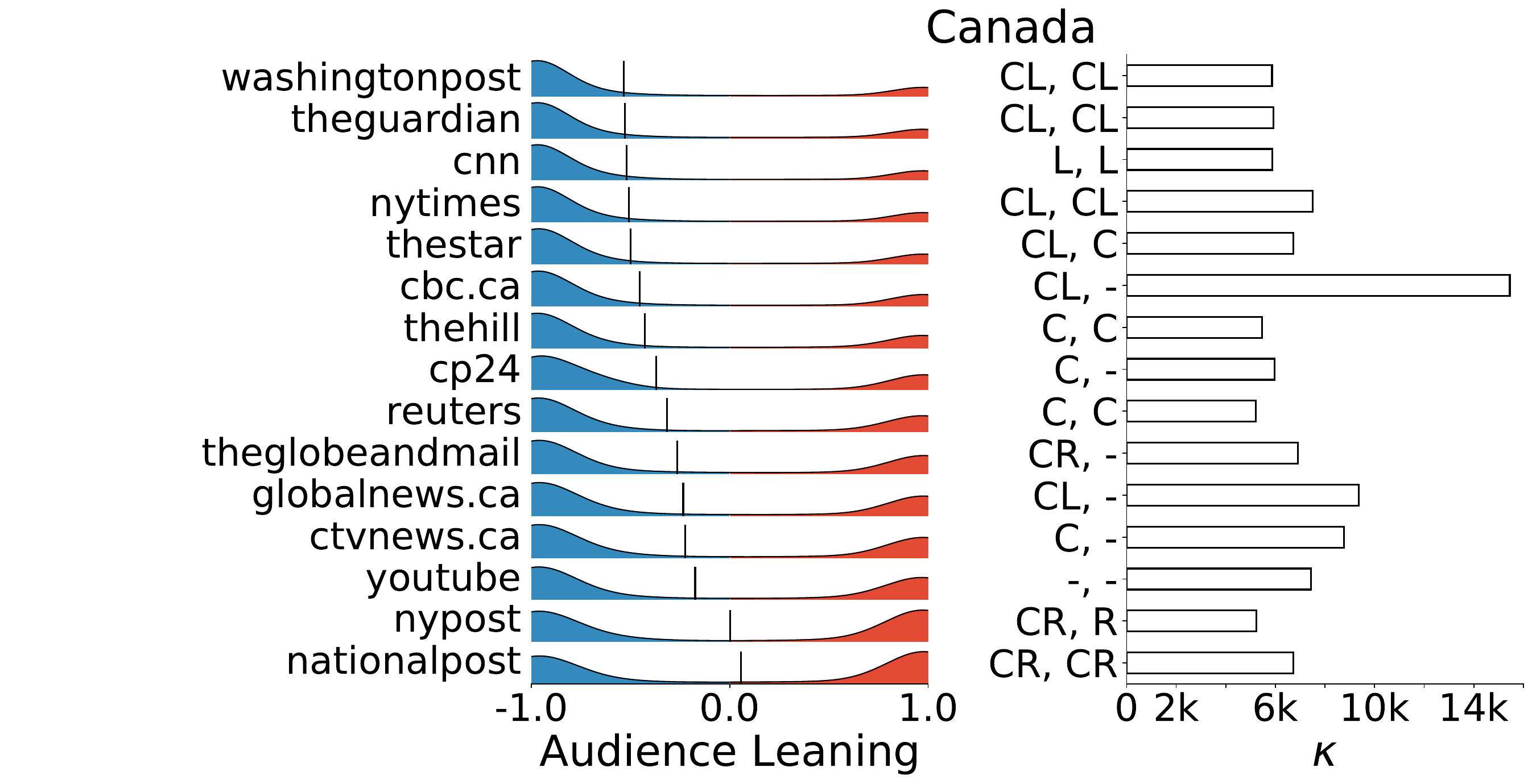}}
    \subcaptionbox{}[.49\columnwidth][c]{%
    \includegraphics[width=.49\columnwidth]{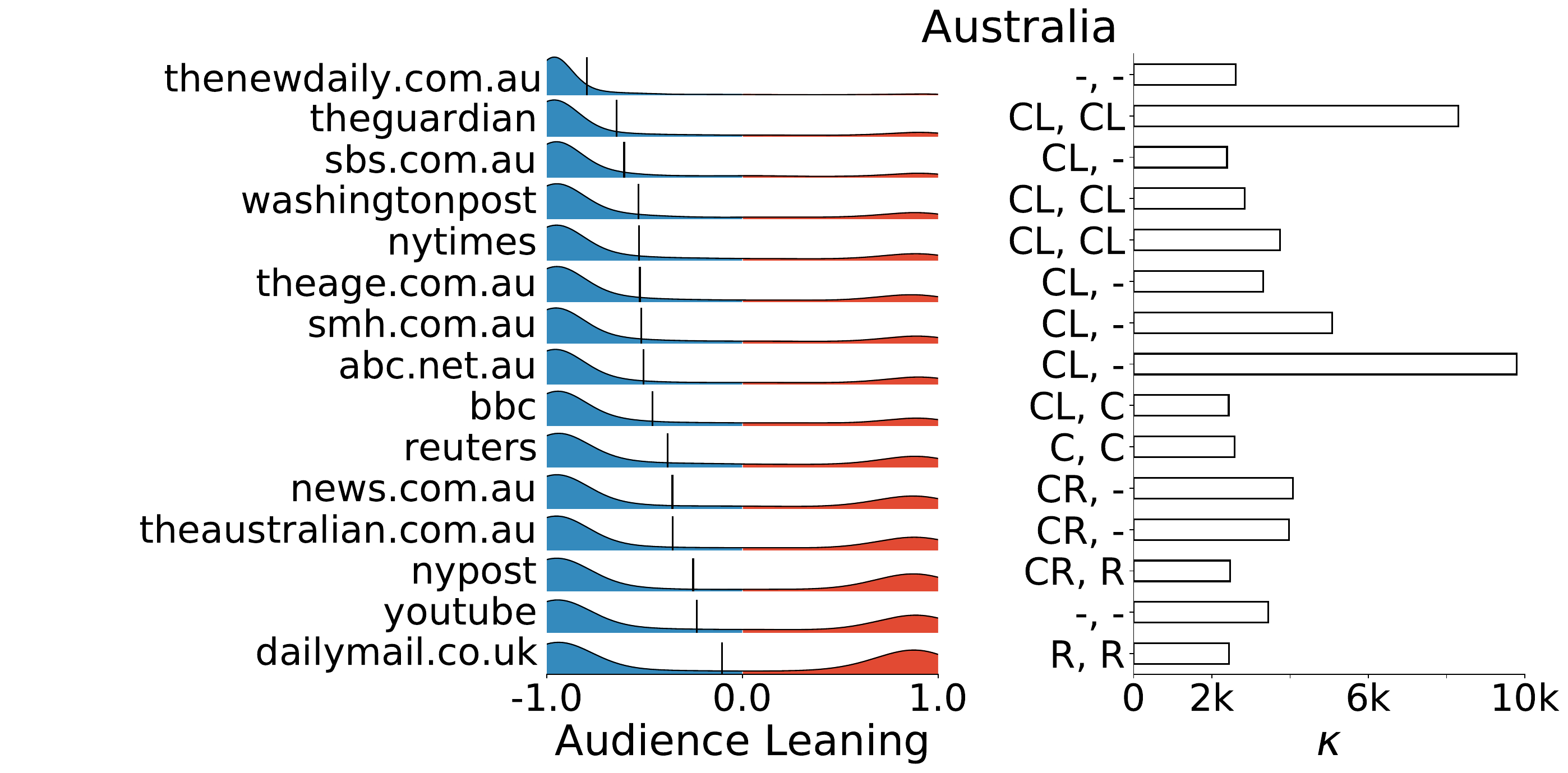}}
    \subcaptionbox{}[.49\columnwidth][c]{%
    \includegraphics[width=.49\columnwidth]{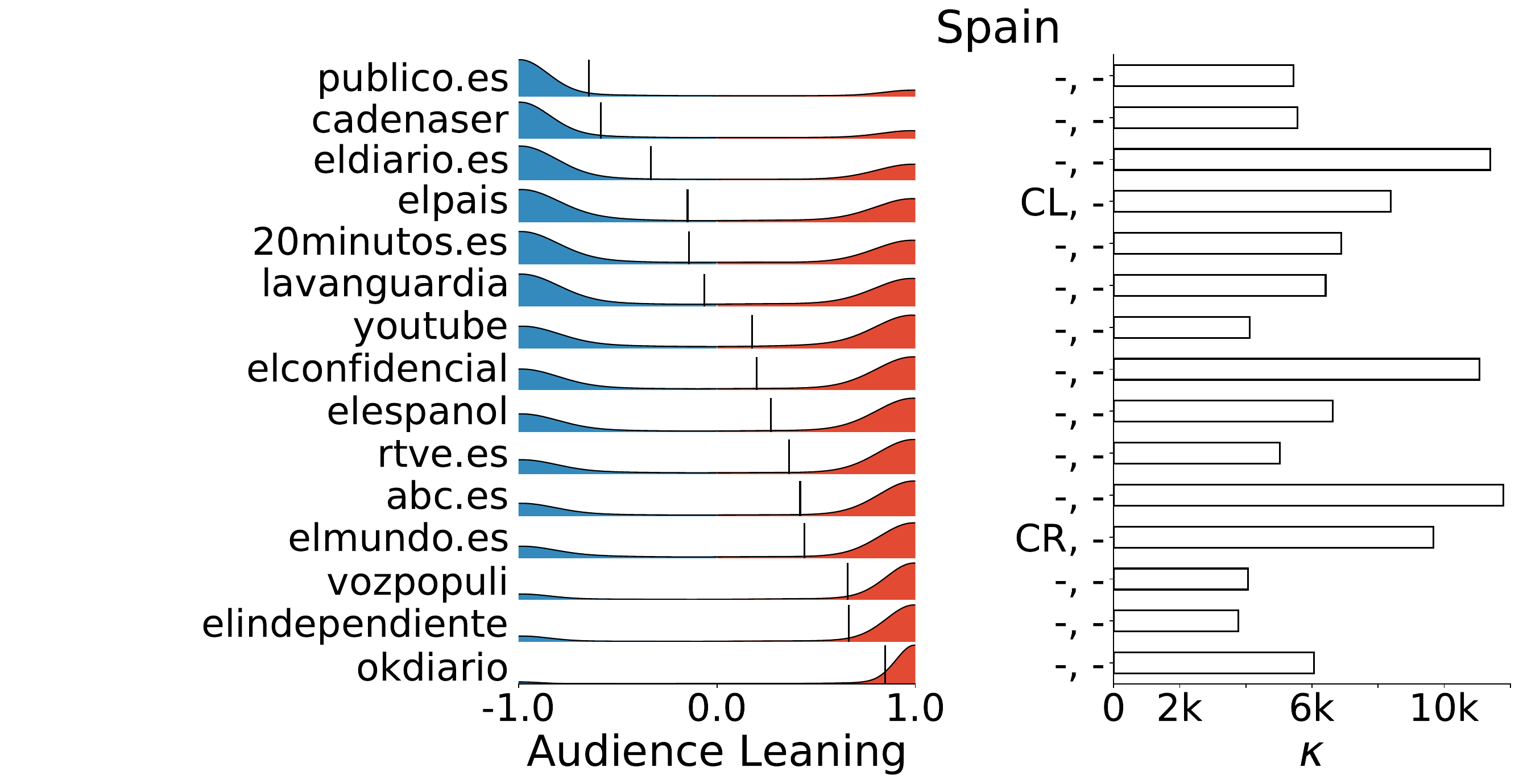}}
    \subcaptionbox{}[.49\columnwidth][c]{%
    \includegraphics[width=.49\columnwidth]{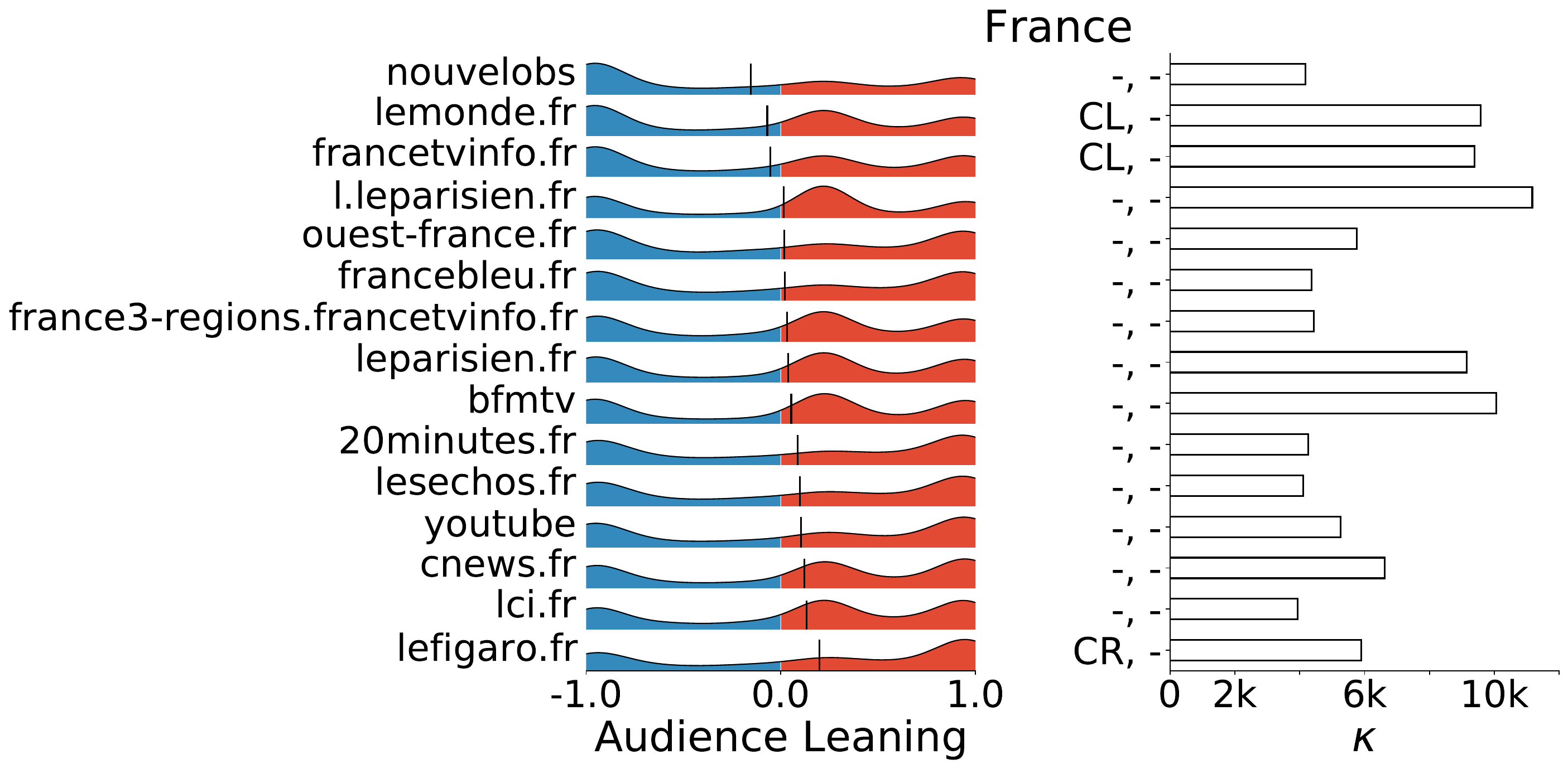}}
    \subcaptionbox{}[.49\columnwidth][c]{%
    \includegraphics[width=.49\columnwidth]{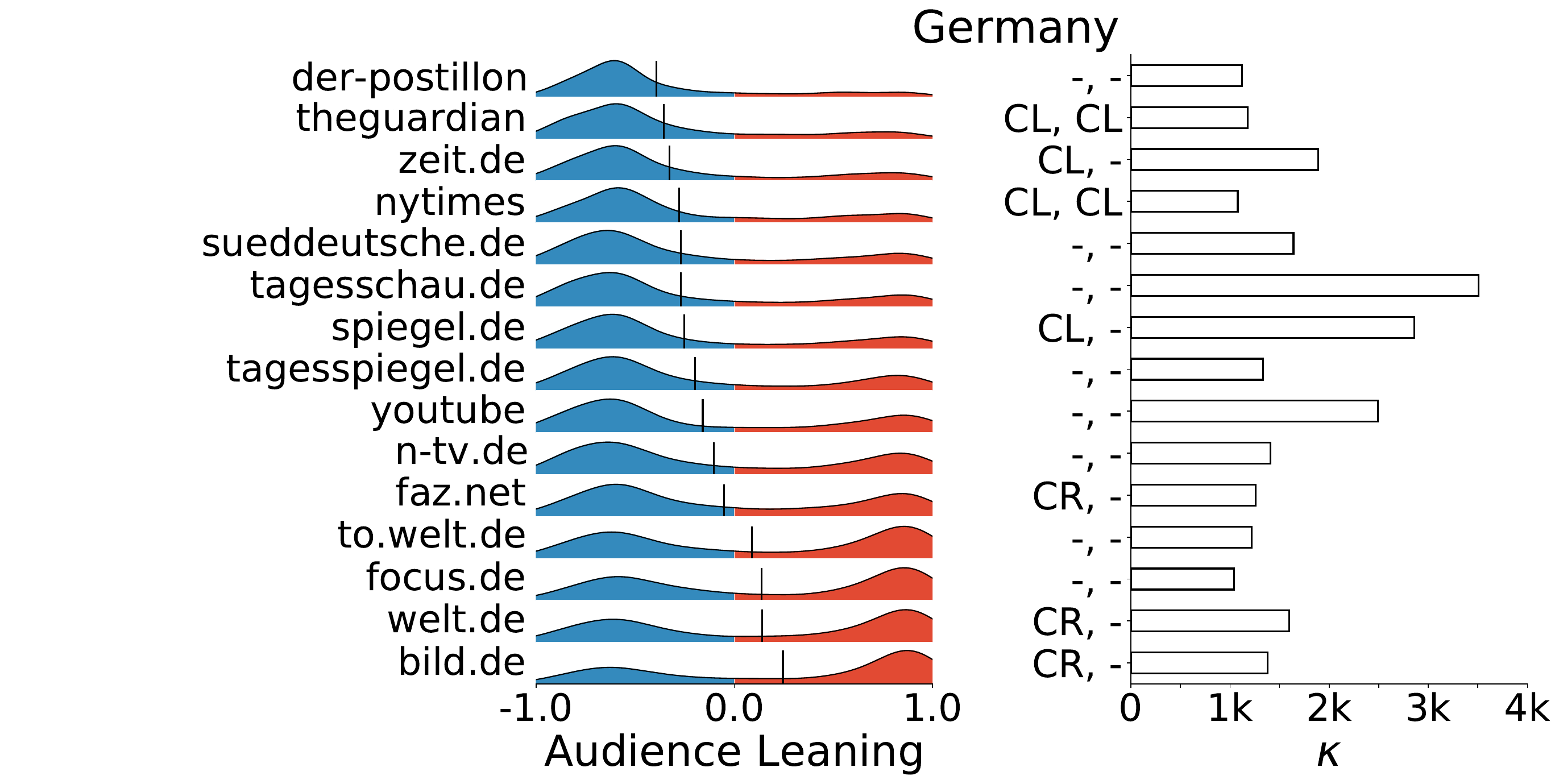}}
    \subcaptionbox{}[.49\columnwidth][c]{%
    \includegraphics[width=.49\columnwidth]{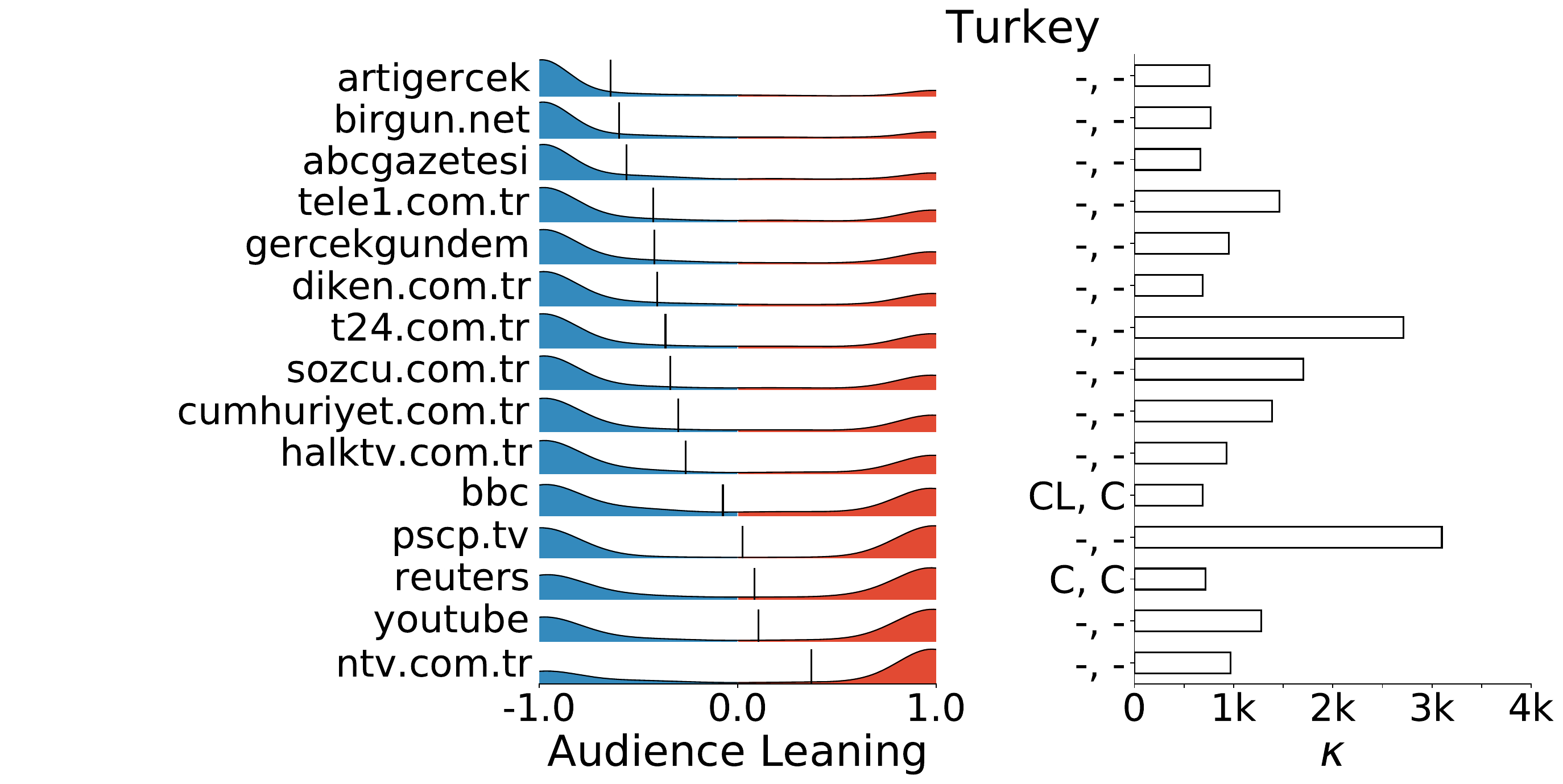}}
    \caption{Audience leaning distribution for the top 15 media domains ranked by audience reach in the eight selected countries. In each subfigure, left part: ridge plots of audience leaning distribution; black vertical line: mean. The domains are ranked by their average audience leaning scores $\bar p(d, l)$ (solid lines). Media at the top (bottom) have more left-leaning (right-leaning) audience.
    Right part: bar plots of audience reach $\kappa(d, l)$ of each domain in the respective country. x-axis: audience reach in thousands. y-axis: media bias ratings from \citet{mbfc} and \citet{allsides}, ``-'' means not rated by the corresponding source.}
    \label{fig:top15-domain}
\end{figure*}

\header{A country-centric view.} 
\Cref{fig:top15-domain} profiles the distribution of audience leaning scores for the top 15 media domains by within-country audience reach in the eight selected countries.\footnote{Note that {\it bbc.co.uk} and {\it bbc} both appear in the UK profiles. The former is for browser requests sent from the UK, while the latter is for those sent from other countries. We treat them as two separate domains to be consistent with prior work such as \citet{Bakshy2015ExposureTI, Budak2016FairAB, robertson2018auditing}.} The media domains within each country are then reordered by their average leaning score $\bar p(d, l)$ (shown as a black vertical line). We also show media bias labels from \citet{allsides} and \citet{mbfc} on the legend, wherever labels are available (L: left, CL: center-left, C: center, CR: center-right, R: right, ER: extreme right). Comparing the two editorially curated ratings, MBFC covers more news outlets, especially outside of the US. At a glance, the geolocated users in the \dataset dataset seem to heavily consume news media from the respective country. There is a modest level of news circulation among the four English-speaking countries (e.g., {\it theguardian, washingtonpost, nytimes} appear in the US, UK, Canada, and Australia). In the four non-English-speaking countries, the most popular media are strongly language-specific -- almost all are from the respective country, except for {\it youtube} that appears in all countries.

\begin{figure*}[tbp]
  \centering
  \captionsetup[subfigure]{labelformat=empty}
  \subcaptionbox{}[.49\columnwidth][c]{%
    \includegraphics[width=.49\columnwidth]{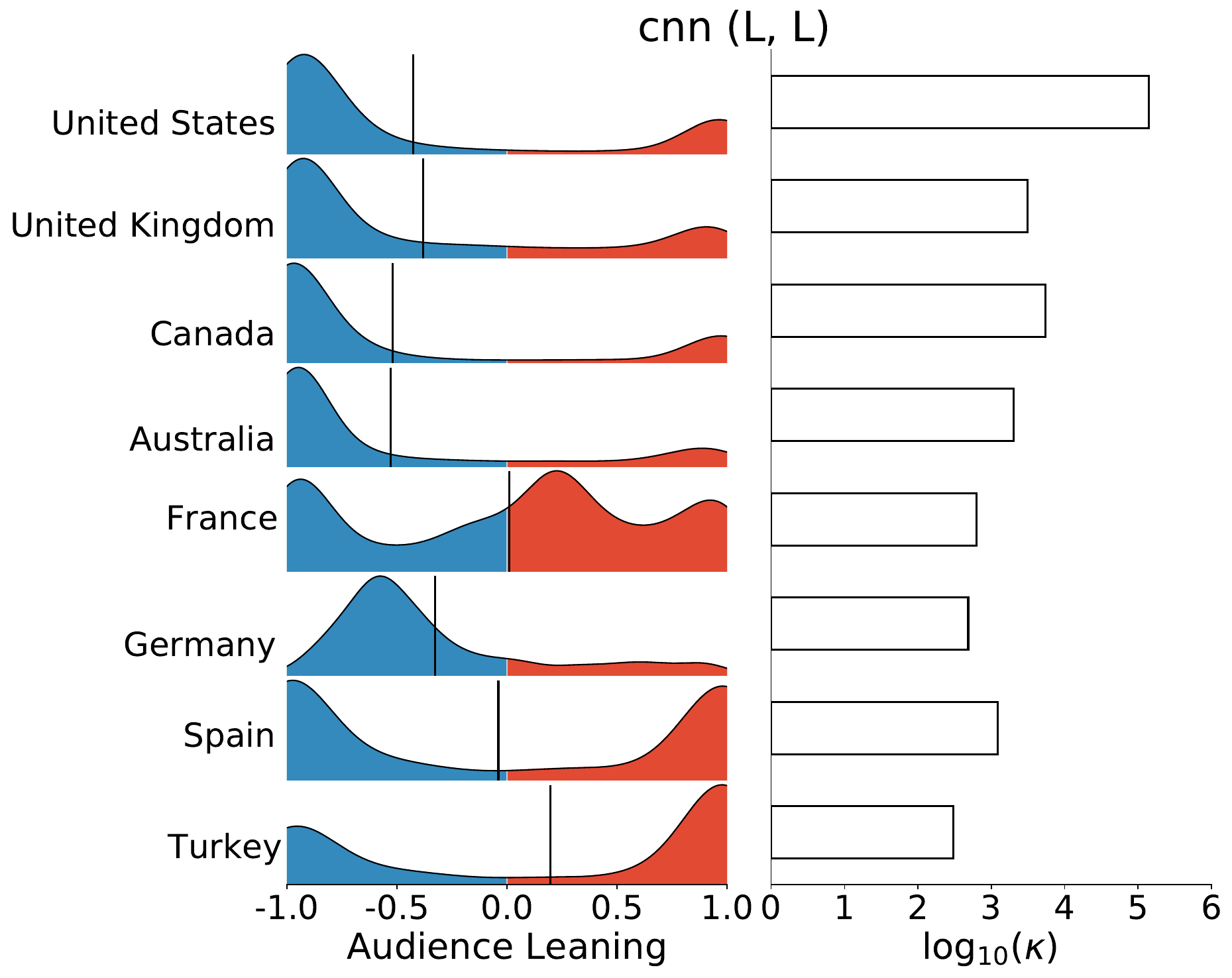}}
  \subcaptionbox{}[.49\columnwidth][c]{%
    \includegraphics[width=.49\columnwidth]{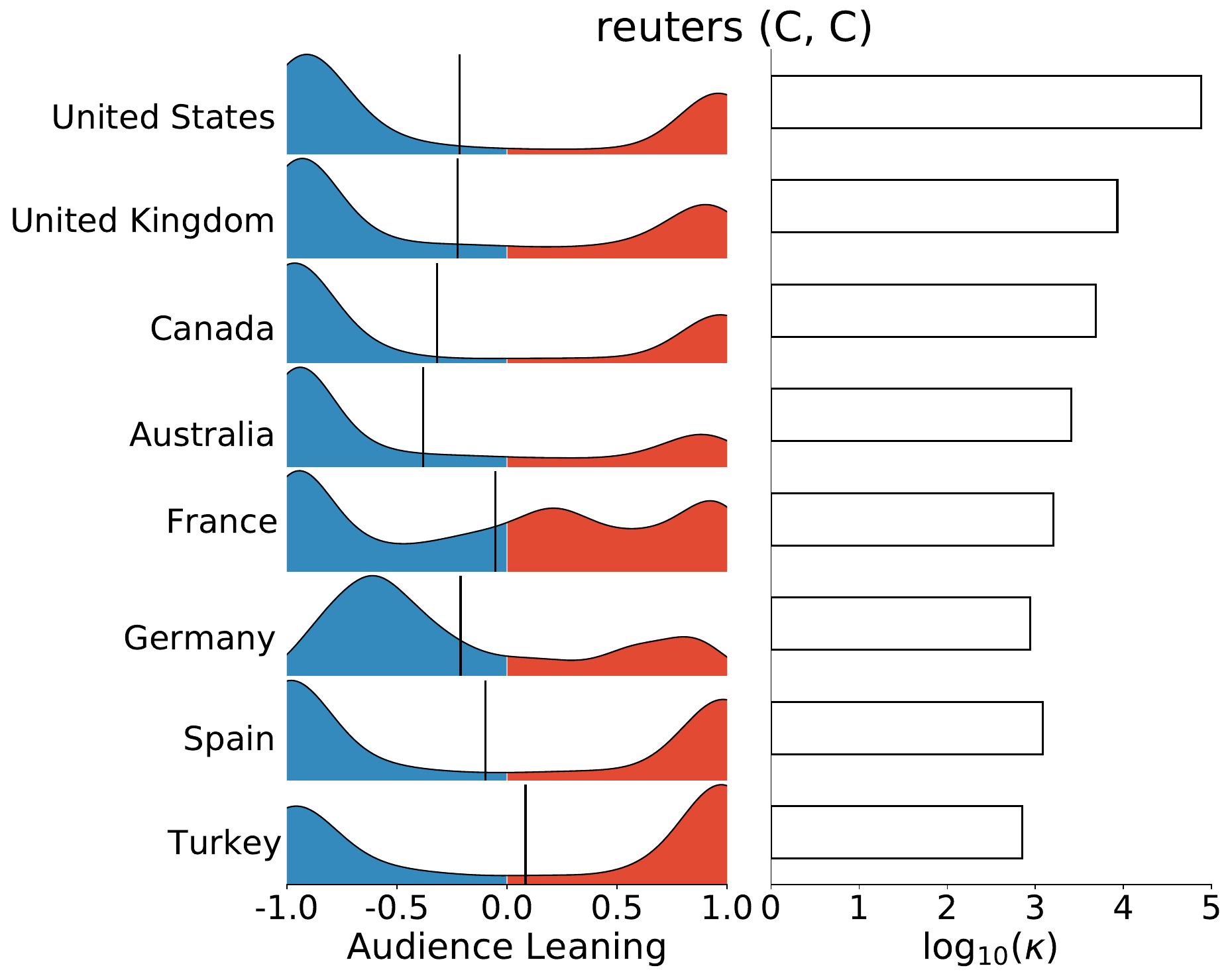}}
  \subcaptionbox{}[.49\columnwidth][c]{%
    \includegraphics[width=.49\columnwidth]{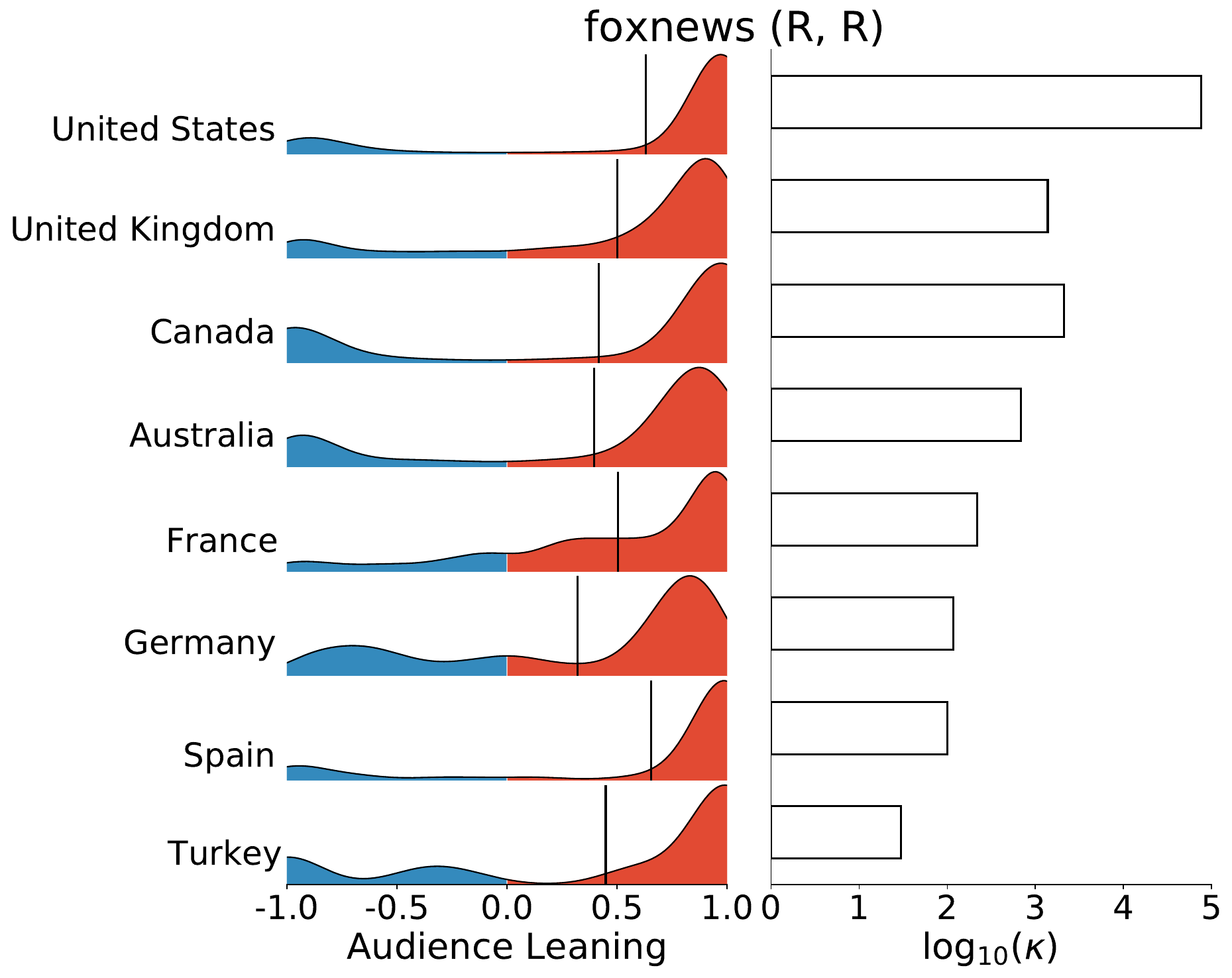}}
 \subcaptionbox{}[.49\columnwidth][c]{%
    \includegraphics[width=.49\columnwidth]{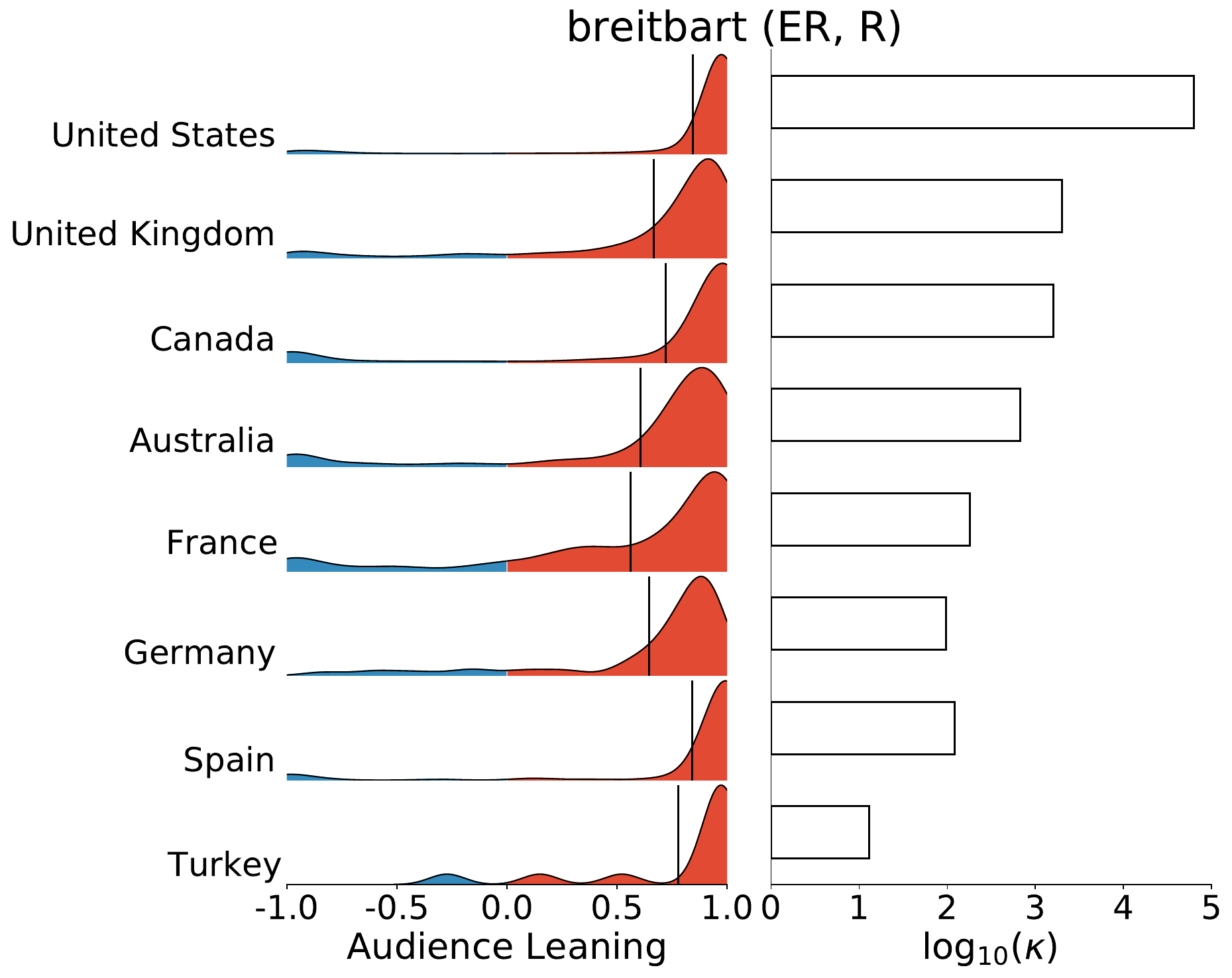}}
    \caption{Audience distributions across different countries for \textit{cnn}, \textit{reuters}, \textit{foxnews}, and \textit{breitbart}. In each subfigure, left part: ridge plots of audience leaning distribution in eight countries; black vertical line: mean. Right part: bar plots of audience reach in each country in log scale. x-axis: audience size in thousands. \citet{mbfc} and \citet{allsides} labels are in brackets after each domain name. 
    }
  \label{fig:domain-across-countries}
\end{figure*}

From these distributions, we observe considerable variations in terms of both the mean and spread of audience leanings. This paints a unique picture of the media landscape. For example, firstly, 11 out of the top 15 domains in the US are either center or left-leaning if we only look at the labels from MBFC and AllSides. However, there exist large variances in their audience distribution -- audiences of domains such as {\it nbcnews, politico, theguardian} consist of a large proportion of left-leaning users in the US, while domains like {\it newsweek, cbsnews, reuters} are shown to have a more balanced audience base. 
Secondly, as the media becomes more partisan, its audience base shrinks to a narrower range. For example, {\it reuters} in the UK consists of a balanced set of left-leaning and right-leaning users, but partisan media {\it mirror.co.uk} and {\it dailymail.co.uk} have few audience from the opposite political side. For far-right media such as {\it breitbart} in the US, {\it okdiario} in Spain, {\it ntv.com.tr} in Turkey, almost no left-leaning users consume it. All this information is not readily available without profiling the distribution of audience leaning scores.
Thirdly, the data-driven leaning estimates can enrich or contradict existing media bias labels
For example, {\it cnn} is rated left-leaning by both sources, but 
the average audience leaning of {\it nbcnews} and {\it washingtonpost}, both rated center-left, are more left (smaller) than that of {\it cnn}.
Fourth, in countries other than the US, the relative audience leaning for media can confirm or contradict default intuitions. In the UK, most media has audiences that are spread across the left-right spectrum, including {\it bbc} (a government-sponsored media) and {\it www.gov.uk} (the government website itself). In Canada, {\it theguardian} ranks more right than {\it washingtonpost}, but it's the opposite in the US.
Lastly, this audience leaning distribution provides a new angle to estimate the biases for media not collected in sources like MBFC or AllSides, or from understudied non-English-speaking countries. For example, we can generate media bias estimates for 12 unrated domains in Spain (less youtube, same for the other countries), 11 in France, 7 in Germany, and 12 in Turkey. In Australia, all media outlets have an average audience leaning score less than 0 (lean left); this is certainly influenced by the background distribution of Australian Twitter users in \dataset (71\% left). Such observationx indicate that the mean audience leaning score is more readily interpreted relative to each other rather than in an absolute sense, and that background statistics in the sample population are important.

\header{A media-centric view.}
\Cref{fig:domain-across-countries} profiles the distribution of audience leaning scores for four example media across the eight countries. We choose {\it cnn}, {\it reuters}, {\it foxnews} and {\it breitbart}, which have been labeled as left, center, right, and extremely-right by~\citet{mbfc}. For each media, the computed average audience leaning scores have a large variance across countries (i.e., audience bases).
For instance, {\it cnn} is shown to have average audience leaning scores varying between -0.5 to 0.2. Despite being rated as left-leaning, the calculation over different countries reveals a more nuanced picture -- it has an average audience leaning score close to zero in France and Spain and even a positive score in Turkey. It has a bimodal distribution in six countries, except for a unimodal distribution in Germany and a trimodal distribution in France.
For center media {\it reuters}, its average audience leaning is close to 0 in France, Spain and Turkey, but most of its audience in Canada, Australia, and Germany are estimated left-leaning. Turkey shows an opposite observation, where about half of the audience are estimated to be right-leaning. In most countries, the distributions are bimodal except for France and Germany.
For right-leaning media {\it foxnews}, the average leaning scores vary from 0.32 to 0.65 across the selected countries. Its audience has been shown to be mainly right-leaning users in the US, UK, France, and Spain (> 80\%); while there is also a nontrivial number of audience occupying the left-leaning spectrum (> 25\%) in Canada, Australia, Germany, and Turkey.
The far-right media {\it breitbart} has average audience leaning scores ranging from 0.5 to 0.85 in the eight countries. Audience leaning distributions of {\it breitbart} are unimodal in all countries, suggesting that very few left-leaning users would consume {\it breitbart} news. Its audiences in the US, UK, Spain, and Turkey are over 90\% right-leaning, whereas in Canada, Australia, France, and Germany, less than 90\% but still more than 75\% of its audiences are estimated right-leaning. 

Our observations are significant due to two reasons. Firstly, while there are many editor-curated~\cite{allsides,mbfc} and data-driven~\cite{Bakshy2015ExposureTI,Budak2016FairAB,robertson2018auditing, fletcher2020polarized} media bias estimates, all of them focus on a notion of \textit{average} leaning, whereas in this work we quantify the spread of the audience base. Secondly, our data-driven measure reveals significant cross-country variation in media domains, the average audience leanings, as well as the political diversity in the audience base.

\section{Conclusion and Discussion}
\label{sec:conclusion}

In this work, we compute a set of measurements to profile the audience of media outlets in eight countries. In particular, our profiling describes not just the average, but also the spread of user leanings for a given media outlet in a given country. Components of our quantitative method include: a new \covid dataset with high coverage; a highly accurate set of geolocated users based on their geotagged tweets and Twitter profiles; a set of comparatively estimated political leaning scores across countries; and a comprehensive validation of the audience leaning score against existing media bias reporting. We use the distributions of audience leaning to profile different media outlets within a specific country, and also profile the audience of a given outlet across different countries. 

\header{Broader implications.}
We posit that these new measures will provide scholars in political science and communications with a data-driven view of a diverse set of countries. The observations can inform media outlets of their comparative effectiveness within and across countries, and give activist groups and individuals insights with which to reflect on our day-to-day media diet. We posit that the methodology used here is transferrable to other large datasets from Twitter and other domains.

\header{Ethical considerations.}
In this work, we report aggregate statistics and do not publish extracted geolocation, nor predicted political leaning of any individual Twitter users. We release datasets according to Twitter's terms of service and guidelines to academic researchers.
We caution that interpretations of media leaning scores are influenced by audience sample, background distribution of left vs. right-leaning users in any country, and the importance of including non-English query terms in work that applies this method to other domains (see limitations below).

\header{Limitations.} It is important to acknowledge a number of limitations related to the data collection and filtering, user leaning estimation, and result interpretation.

\begin{itemize}[leftmargin=*]
\item Despite \covid being the dominant topic over the world in 2020, the media leanings and audience base are only estimated from \covid related content in this work, rather than a representative set of all media content. An alternative data collection strategy would be extracting all news articles from the 1\% Twitter data stream, and then replicating the geolocation extraction and political leaning estimation steps from this work. Much prior research studies the general news sharing on social platforms~\cite{an2014partisan,schmidt2017anatomy}. However, one potential risk of using the general news articles is the topic misalignment of public discourse across different countries. Anchoring on a specific topic ensures the maximal comparability.
\item The choices of English-only or English-majority query terms and hashtags introduce bias in the subsequent content analysis for non-English speaking countries. This is a gap that the current author team does not have the expertise to bridge. Countries such as Japan, South Korea, and Thailand have a significant Twitter presence but are not included due to the \covid related queries being in the Latin alphabet and will not match their respective native languages. 
\item The profiled audience bases may still bias toward the subset of English-speaking Twitter users in the four non-English-speaking countries.
\item This method only applies to countries with a significant Twitter presence. Users in many countries congregate on locally-preferred social media platforms (such as WhatsApp, Viber, Naver, among a large number of others). This means that our observations are limited to those who also engage on Twitter, despite that our proposed methodology could still be applied to data from other platforms, other countries, or other topics. Furthermore, user representation in countries with tight state censors is necessarily skewed towards those with the willingness and technical means to circumvent such barriers. 
\item While the retweet interactions with local politicians surface a collection of politically divided users, there is still an unknown number of politically modest users left out from this study.
\item The \dataset tweet dataset focuses on \covid related topics and is collected during a period when \covid attracts an unprecedented amount of attention all over the world. We are therefore cautious about generalizing our findings to the general media consumption patterns or to other topics. Because if we replace the \covid with another topic (e.g., abortion rights, gun control, or Black Lives Matter activism~\cite{lee2022whose}), we will obtain a new set of estimated user political leaning scores, thus a new set of media bias estimates. Some observations in the current manuscript may change under other topics.
\end{itemize}

The \covid pandemic is a time when the whole world has come together to fight a global crisis. Given the \covid news consumption can influence public health behaviors (e.g., social distancing, mask wearing, vaccine belief), it is of great importance to understand the news consumption patterns during the pandemic. This quantitative analysis presented in this work should be considered a supplement to, rather than a replacement for, in-depth examinations of media systems and dynamics in different countries. The analysis could be interpreted alongside data and analysis from other sources, such as politicians' Twitter presence and interactions~\cite{huszar2022algorithmic}. We hope this work could lay the foundation for future analysis of media ecosystems in times of disaster.

\header{Future work.} It will be interesting to intersect media audience leanings with the public pandemic attitudes and actions in different countries. One could also quantify the robustness of media bias metrics with respect to different (and likely lower) sampling rates of Twitter data. We desire to expand Twitter queries and the scope of political leaning estimates, to potentially have multi-dimensional descriptions of media audiences over a larger number of countries. Another methodology challenge deserving more community efforts is the political leaning estimation on high-dimensional data, which can shift the research focus from modeling the oversimplified one-dimensional political axis to the two-dimensional Socio-Economic political compass.

\section*{Acknowledgments}
This work is supported in part by AFOSR Grant FA2386-20-1-4064. We are grateful for the computing and infrastructure support by Nectar Research Cloud, a collaborative Australian research platform supported by the NCRIS-funded Australian Research Data Commons (ARDC). We also thank the anonymous reviewers and area chairs for their valuable comments that helped shape our methodology and results. 

\bibliographystyle{ACM-Reference-Format}
\bibliography{main}

\appendix

\section{COVID-19 Keywords and Crawler Settings}
\label{app:keywords}

We used 12 Twitter sub-streams to collect the \covid tweets. The settings and assigned keywords are displayed below in JSON format. The \texttt{language} attribute indicates the corresponding collected language of tweets, with empty list indicating all languages.

\header{All keyword list.} [``coronavirus'', ``covid19'', ``covid'', ``covid–19'', ``COVID---19'', ``pandemic'', ``covd'', ``ncov'', ``corona'', ``corona virus'', ``sars-cov-2'', ``sarscov2'', ``koronavirus'', ``wuhancoronavirus'', ``wuhanvirus'', ``wuhan virus'', ``chinese virus'', ``chinesevirus'', ``china'', ``wuhanlockdown'', ``wuhan'', ``kungflu'', ``sinophobia'', ``n95'', ``world health organization'', ``cdc'', ``outbreak'', ``epidemic'', ``lockdown'', ``panic buying'', ``panicbuying'', ``socialdistance'', ``social distance'', ``socialdistancing'', ``social distancing'']

\header{sub-stream 1}
\begin{itemize}[leftmargin=*]
    \item keywords: [``wuhanlockdown'', ``wuhan'', ``kungflu'', ``sinophobia'', ``n95'', ``world health organization'', ``cdc'', ``outbreak'', ``epidemic'']
    \item languages: []
\end{itemize}

\header{sub-stream 2}
\begin{itemize}[leftmargin=*]
    \item keywords: [``lockdown'', ``panic buying'', ``panicbuying'', ``socialdistance'', ``social distance'', ``socialdistancing'', ``social distancing'']
    \item languages: []
\end{itemize}

\header{sub-stream 3}
\begin{itemize}[leftmargin=*]
    \item keywords: [``pandemic'', ``covd'', ``ncov''],
    \item languages: [``en'', ``es'']
\end{itemize}

\header{sub-stream 4}
\begin{itemize}[leftmargin=*]
    \item keywords: [``coronavirus'']
    \item languages: [``en'']
\end{itemize}

\header{sub-stream 5}
\begin{itemize}[leftmargin=*]
    \item keywords: [``covid'']
    \item languages: [``en'']
\end{itemize}

\header{sub-stream 6}
\begin{itemize}[leftmargin=*]
    \item keywords: [``covid–19'', ``COVID---19'', ``covid19'']
    \item languages: [``en'']
\end{itemize}

\header{sub-stream 7}
\begin{itemize}[leftmargin=*]
    \item keywords: [``corona'', ``corona virus'', ``sars-cov-2'', ``sarscov2'', ``koronavirus'', ``wuhancoronavirus'', ``wuhanvirus'', ``wuhan virus'', ``chinese virus'', ``chinesevirus'', ``china'']
    \item languages: [``en'']
\end{itemize}

\header{sub-stream 8}
\begin{itemize}[leftmargin=*]
    \item keywords: [``coronavirus'', ``corona'', ``corona virus'']
    \item languages: [``es'']
\end{itemize}

\header{sub-stream 9}
\begin{itemize}[leftmargin=*]
    \item keywords: [``covid19'', ``covid'', ``covid–19'', ``COVID---19'', ``sars-cov-2'', ``sarscov2'', ``koronavirus'', ``wuhancoronavirus'', ``wuhanvirus'', ``wuhan virus'', ``chinese virus'', ``chinesevirus'', ``china''],
    \item languages: [``es'']
\end{itemize}

\header{sub-stream 10}
\begin{itemize}[leftmargin=*]
    \item keywords: [``coronavirus'', ``corona'', ``corona virus'', ``covid19'', ``covid'', ``covid–19'', ``COVID---19'', ``sars-cov-2'', ``sarscov2'', ``koronavirus'', ``wuhancoronavirus'', ``wuhanvirus'', ``wuhan virus'', ``chinese virus'', ``chinesevirus'', ``china'']
    \item languages: [``th'', ``it'', ``fr'', ``tr'']
\end{itemize}

\header{sub-stream 11}
\begin{itemize}[leftmargin=*]
    \item keywords: [``coronavirus'', ``corona'', ``corona virus'', ``covid19'', ``covid'', ``covid–19'', ``COVID---19'', ``sars-cov-2'', ``sarscov2'', ``koronavirus'', ``wuhancoronavirus'', ``wuhanvirus'', ``wuhan virus'', ``chinese virus'', ``chinesevirus'', ``china'']
    \item languages: [``in'', ``ko'', ``ja'', ``und'']
\end{itemize}

\header{sub-stream 12}
\begin{itemize}[leftmargin=*]
    \item keywords: [``coronavirus'', ``corona'', ``corona virus'', ``covid19'', ``covid'', ``covid–19'', ``COVID---19'', ``sars-cov-2'', ``sarscov2'', ``koronavirus'', ``wuhancoronavirus'', ``wuhanvirus'', ``wuhan virus'', ``chinese virus'', ``chinesevirus'', ``china'']
    \item languages: [``pt'', ``zh'', ``ar'', ``de'', ``tl'', ``cs'', ``vi'', ``pl'', ``ru'', ``sr'', ``el'', ``nl'', ``hi'', ``da'', ``ro'', ``is'', ``no'', ``hu'', ``fi'', ``lv'', ``et'', ``bg'', ``ht'', ``uk'', ``lt'', ``cy'', ``ka'', ``ur'', ``sv'', ``ta'', ``sl'', ``iw'', ``ne'', ``fa'', ``am'', ``te'', ``km'', ``ckb'', ``hy'', ``eu'', ``bn'', ``si'', ``my'', ``pa'', ``ml'', ``gu'', ``kn'', ``ps'', ``mr'', ``sd'', ``lo'', ``or'', ``bo'', ``ug'', ``dv'', ``ca'']
\end{itemize}

\section{Political Dividing Hashtags for the 2020 US Presidential Election}
\label{app:politics-hashtags}

\begin{table*}[b]
  \centering
  \small
  \begin{tabular}{p{13.56cm}}
    \toprule
     cult45sucks, ditchmitch, lockuptrump, trumppandemic, thanksobama, impeach, joebiden2020, stevencrowderdrinksdogcum, byedon2020, impeachthemf, trumpisdone, votebiden, leftwing, bunkerbabybonespurs, removetrump, dumptrumphesachump, 45notmypresident, 25thamendment, trumpisatrator, wrongtrump, wearredvoteblue, fuckdonaldtrump, rumpresident, bernietulsi2020, covidiots, uniteblue, medicare4all, noantiblackracism, turnthesenateblue, yanggang2020, onlybernie, americaneedsyang, cult45, yangorbust, shit\_trump\_would\_say, joebiden2020forpresident, nevertrump, faketrumpf17klies, anyonebuttrump, idiottrump, trumpgonnagetschooled, firetheliar2020, pencedemic, trumpsupportersneedtobekilled, warren2020, taketheorangeclownout2020, settleforbiden, stillvotingyang, moretrumplies, prochoiceoneverthingoneverthing, traitors4trump, thetrumpvirus, trumptrash, votebernie, bidenharris2020touniteandrebuildamerica, cowdercrowder, prochoice, republicansforjoebiden, putinspuppet, outtrump, womenforbernie, fuckthegop, trumpsucks, bluetsunami, bunkerbaby, bluewave, liberal, trumpliespeopledie, impeachtrump, trumpvirus2020, trumpforprison2020, leftisbest, voteagainsttrump, typhoidtrump, biden2020, whiteprivilegeisreal, trumpshutdowncps, imwithher, trumptarded, trumpisthehoax, bernieorvest, trumpliesaboutcoronavirus, berniecomeback, trumpexposed, crimesagainsthumanity, obamaenvy, yanggangforever, berniebeatstrump, bidenforpresident, trumpcult, theresistance, yangmediablackout, resistance, voteprogressives, votebiden2020, tre45on, lockhimup, gojoe, trump4prison2020, trumpisnotwell, trumpforprison, trumpplague, traitortrump, neverdonaldtrump, trumpimpeachment, magaisformorons, voteblue2020, byedon, shoottrumpnow, fucktrump, soundslikeyang, donaldwho, bernieismypresident, votebluetoendthisnightmare, trumpslump, trumptard, resisttrump, bernieorbust, orangemoron, trumpresign, dumptrump2020, berniesanders2020, medicareforall, itrustbernie, trumplies, couldahadyang, feelthebern2020, joe2020, impeachdrumf, dumpthetrump, cult45goppathetic, trumpmafia, humanityfirst, obamagreat, liberatethewhitehouse, writeinbernie, gopprodeath, donthecon, trumpflu, impeachbarr, blackwomenforyang, usnotme, trump4prisondent2020, ilikebernie, presidentialfailure, impotus, fakepotus, russiagate, byetrump, bernie2020, cult45gop, yang2020, bluenomatterwho, notmypresident, impotus45, ripgop, firetrump, resist, nevadaforpresidenttrump, teampelosi, nevertrumper, jailtrump, failedtrump, bluedowntheballot2020, lettulsispeak, dictatordrumpf, lockthemallup, trumpcrash, foreverbernie, drumpf, swampydon, voteblue, republicanpandemic, bunkerboy, votingallblue2020, bidenharris, votebluenomatterwho2020, bernthednc, resigntrump, stopdonaldtrump, republicansagainsttrump, bluewave2020, endtrumpism, andrewyang2020, trumpthecorrupt, votebluenomatterwho, crookeddonald, andrewyang4pres, votethemallout, euthanizetrumpsupporters, throwouttrump, joebiden4president, bluemaga, moscowmitch, treasonoustrump, sanders2016, presidentsanders, cuomo2024, yanggang2024, yangbeatstrump, magavirus, lockthemup, trumpgate, yangganglove, presidentpussyassbitch, votethemout, votetrumpout2020, gothsforbernie, youtubeandrewyang, imwithhilary, votetrumpout, fakepresident, coldfeetcrowder, votebluein2020, sanders2020, onevoicel, trumpdepression, letyangspeak, trumptardforprison2020, makeamericahonestagain, justicedemocrats, diaperdonald, mybodymychoice, dementiadonnie, 1yang2024, time4trumptoburn, yangmentum, downwithnazitrump, bernie2020ournations21stcenturyfdr, traitor45forprison2020, magats, christiansagainsttrump, voteblueinnovember, cuomoforpresident, trumpfuckedup, countryaboveparty, takehimout, yang2024, shameontrump, trumpthemurderer, magatards, presidentberniesanders, votebluetosaveamerica, notmeus, ridinwithbiden, fakepotusrealcriminal, flipthesenate, berniebros, resignnow, republictards, regredvoteblue, trumpandemic, tulsi2020, presidentcuomo, anyonebuttrump2020, locktrumpup, feelthebern, bunkerbitch, stillsanders, americaortrump, standwithtulsi, donaldsucks, impeachdt, votehimout, onlybernie2020, nevertrumpers, stoptrump, bernieorbust2020, trumpsfault, trumppenceoutnow, dumptrump, trumpfraud, donaldjtrumpfebruaryfailure, comradetrump, berniebro, trumpthelyinginsanefascistbigot, yangwillwin, drainthewhitehouse, voteblueacrosstheballot, trumpthetraitor, connecttheleft, resistance2020, liarinchief, bidenridenbluewave, trumpfearsyang, liberal, democrat, democrats, democratic, bidenwins2020, reopentard, despicabledonaldtrump, trumpsupportersareinferior, voteforbernie, fucktrump2020, bernie2016, trumpprison2020, bidenharris2020, yangwasright, cuomo2020, defundtrump, bunkerboytrump, thanksbiden, bluetsunami2020, impeached, trumpcrimefamily, farmersagainsttrump, trumpvirus, trumpviruscoverup, bloomberg2020, republicansareterrorists, bernienina2020, berniecares, m4a, worstpresidentever \\
    \bottomrule
  \end{tabular}
  \caption{Left-leaning hashtags.}
  \label{app:political-hashtag-list-1}
\end{table*}

\begin{table}[t]
  \centering
  \small
  \begin{tabular}{p{13.56cm}}
    \toprule
     walkedaway, blacks4trump2020, supertrump, wwg1wgaww, patriotsforafreeamerica, wakeupamerica, defundthedemocrats, nevervotedemokratagain, magahats, nra, darktolight, greatawakening, voterfraud, patriotsunite, rednationrising, latinos4trump, thinblueline, fourmoreyears, impeachmentmeansnothing, shutdowntheleft, thecollectiveq, donjr2024, trumpnation, qanon, bluethinline, trumpforever, demokkkrat, trumpcheerleaders, teamtrump, kingtrump4ever, trump2020landslidevictory, buildthewall, prolife, pence2020, wwg1wga, kaga2020, drainingtheswamp, godwins, qanonproof, bluelifematters, australiafortrump, trump2020wwg1wga, imwithq, blacksfortrump2020, sayno2vaccines, puertoricans4trump, stupidliberal, hispanicwomen4trump, keepamericagreat, veteransfortrump, demokkkrats, pro2a, magababymaga, democratsaredummies, trump2028, drainthatswamp, womenfortrump, trump2024, kaga, potustrumpmaga, supporttheblue, redpill, lovepresidenttrump, liberalssuck, godwinsalways, qarmy, libtards, defundthedemoncrats, qdigitalarmy, votredtsunami2020, trumppence2020, qanon8chan, democratssuck, trumpusamerica, voteredtosaveamerica, wethepeople, voteredtosaveamerica2020, america1st, ccot, filmyourhospital, gaysfortrump, itsnotyourbody, obamagate, 4mreyrs, trumpisyourpresident, latiosfortrump, blacks4trump, trumpderangementsyndrome, trumptrain2020, lovepotus3545, qanons, kag, lationsfortrump, wedonotconsent, bestpresidentever, americansfirst, maga2020, stillyourpresident, qanonmerch, trump2020, draintheswamp, kag2020, trump4eva, 4moreyears, noblackvoteforbiden, trumpocrat, proudpatriot, maga, votedemsout, uniteright, magarapper, istandwithtrump, qpatriot, bestpresidentever45, lovemypresident, trump4life, trump2q2q, godblessourpresident, bidenishidingbecausehismindissliding, trumpforallallfortrump, redpilldiaries, magabeanie, impeachdem, patriots, 2astrong, lawandorder, witnessingthegreatawakening, mastertrump, americafirst, lockherup, votedemout, godlives, chicanoconservative, trumpourimmortalbeaconoflight, blackvoicesfortrump, walkway, democratsfortrump, votebluetored, thegreatawakening, makeamericagreatagain, wwg1wga\_ww, djt2020, tucker2024, demexitedin2016, vivatrump, conservatism, votetrump, buildthewallandcrimewillfall, glockherup, dethronethedemoncrats, magaland, uncensoredpatriot, gays4trump, redwaverising, trustq, ivanka2024, walkawaycampaign, trump2016, blacksfortrump, impeachpelosi, armyfortrump, jesusmatters, athiestdonaldjtrumpsupportertrumpandus, mrpresidenttrump, screwliberals, backtheblue, shadowgate, standforourflag, filmyourhospitals, democratsaredestroyingamerica, trump2048, presidenttrump, mexicans4trump, 2a, redwave, conservativenews, wwg1wgall, 45isthebest, trillionairesfortrump, kag2q2q, votetrump2020, candaceowens2024, wwg1wgaworldwide, demonrats, shapiro2024, pray4djt, redpillhardcore, votered, abolishdemocrats, trumptrain, clintonbodycount, conservative, latinos4trump2020, pro2ndamendment, trumpismypresident, qalert, patheticdemocrats, redpilltheyouth, screwthedemocraticparty, gotrump, realq, patriot, bluelivesmatter, redwave2020, liberalismiscancer, conservatism, conservative, republican, republicans, qed, trumplandslide2020, awakening, altright, redpilltaken, votedemsoutofoffice, trumpgirlamericafirst, trumpwwg1wga, keepamericagreatwithfourmoreyears, backthepolice, staywoke, whenyoutakeredpills, wherewegoonewegoall, wakeupisrael, liberalismisamentaldisorder, latinosfortrump, wwg1wwa, mypresident, democratshateamerica, alexjonesisright, tcot, trumpsupporter, wwg1wga\_worldwide, redpillhardcoreradioshow, democratscreatedthekkk, abortionismurder, redpilled, avidtrumpsupporter, wwg1wgalllllll, trumpjr2024, latinosfortrump2020, walkaway, buildthatwall, godisreal, wgw1wga, clintonsforprison, trump2020landslide, 2ndamendment, trump2020nowmorethanever \\
    \bottomrule
  \end{tabular}
  \caption{Right-leaning hashtags.}
  \label{app:political-hashtag-list-2}
\end{table}

\received{January 2023}
\received[revised]{April 2023}
\received[accepted]{July 2023}

\end{document}